\documentclass[aps,pra,superscriptaddress,twocolumn,notitlepage]{revtex4-1}
\usepackage{float}
\usepackage{graphicx}
\usepackage[most,breakable]{tcolorbox}
\usepackage{hyperref}
\usepackage{amsthm,amsmath,amssymb,amsbsy}
\usepackage{colortbl}
\usepackage{physics}
\usepackage{ytableau}
\usepackage{bm}
\usepackage{newtxmath}

\usepackage{booktabs}
    \newcommand\btrule[1]{\specialrule{#1}{0pt}{0pt}}

\newcommand{\proj}[1]{\ketbra{#1}{#1}}
\newcommand\braXket[4][]{\mathinner{\langle#2\vert#3\vert#4\rangle}_{#1}}
\newcommand{\mc}[1]{\mathcal{#1}}
\newcommand{\bg}[1]{\boldsymbol{#1}}
\newcommand{\ox}{\otimes}
\newcommand{\DS}[1]{\Delta S_\text{#1}}

\makeatletter
\newcommand{\thickhline}{
    \noalign {\ifnum 0=`}\fi \hrule height 1pt
    \futurelet \reserved@a \@xhline
}
\newcolumntype{"}{@{\hskip\tabcolsep\vrule width 1pt\hskip\tabcolsep}}
\makeatother

\newtcolorbox[auto counter]{tableBox}[2][]{%
float*=t,width=\textwidth, standard,fonttitle=\bfseries,title=Table~\thetcbcounter: #2,#1,
colback=white,colframe=black!20!orange}

  \newcommand{\inlineheading}[1]{\noindent \textbf{{#1.}}}

\begin{document} 

\title{\large Mixing indistinguishable systems leads to a quantum Gibbs paradox}

\author{Benjamin Yadin}
    \email{benjamin.yadin@physics.ox.ac.uk}
    \affiliation{School of Mathematical Sciences and Centre for the Mathematics and Theoretical Physics of Quantum Non-Equilibrium Systems,
University of Nottingham, University Park, Nottingham NG7 2RD, United Kingdom}
    \affiliation{Wolfson College, University of Oxford, Linton Road, Oxford OX2 6UD, United Kingdom}
\author{Benjamin Morris}
\email{benjamin.morris@nottingham.ac.uk}
    \affiliation{School of Mathematical Sciences and Centre for the Mathematics and Theoretical Physics of Quantum Non-Equilibrium Systems,
University of Nottingham, University Park, Nottingham NG7 2RD, United Kingdom}
\author{Gerardo Adesso}
    \email{gerardo.adesso@nottingham.ac.uk}
    \affiliation{School of Mathematical Sciences and Centre for the Mathematics and Theoretical Physics of Quantum Non-Equilibrium Systems,
University of Nottingham, University Park, Nottingham NG7 2RD, United Kingdom}

\begin{abstract}
    The classical Gibbs paradox concerns the entropy change upon mixing two gases.
Whether an observer assigns an entropy increase to the process depends on their ability to distinguish the gases.
A resolution is that an ``ignorant'' observer, who cannot distinguish the gases, has no way of extracting work by mixing them.
Moving the thought experiment into the quantum realm, we reveal new and surprising behaviour: the ignorant observer can extract work from mixing different gases, even if the gases cannot be directly distinguished.
Moreover, in the macroscopic limit, the quantum case diverges from the classical ideal gas: as much work can be extracted as if the gases were fully distinguishable.
We show that the ignorant observer assigns more microstates to the system than found by naive counting in semiclassical statistical mechanics.
This demonstrates the importance of accounting for the level of knowledge of an observer, and its implications for genuinely quantum modifications to thermodynamics.
\end{abstract}

\maketitle

\section*{Introduction}
\noindent Despite its phenomenological beginnings, thermodynamics has been inextricably linked throughout the past century with the abstract concept of information.
Such connections have proven essential for solving paradoxes in a variety of thought experiments, notably including Maxwell's demon~\cite{bennett2003notes} and Loschmidt's paradox \cite{holian1987resolution}.
This integration between classical thermodynamics and information is also one of the main motivating factors in extending the theory to the quantum realm, where information held by the observer plays a similarly fundamental role~\cite{binder2018thermodynamics}.

This work is concerned with the transition from classical to quantum thermodynamics in the context of the Gibbs paradox~\cite{gibbs1879equilibrium,levitin1992gibbs,allahverdyan2006explanation}.
This thought experiment considers two gases on either side of a box, separated by a partition and with equal volume and pressure on each side.
If the gases are identical, then the box is already in thermal equilibrium, and nothing changes after removal of the partition.
If the gases are distinct, then they mix and expand to fill the volume independently, approaching thermal equilibrium with a corresponding entropy increase.
The (supposed) paradox can be summarised as follows: what if the gases differ in some unobservable or negligible way -- should we ascribe an entropy increase to the mixing process or not?
This question sits uncomfortably with the view that thermodynamical entropy is an objective physical quantity.

Various resolutions have been described, from phenomenological thermodynamics to statistical mechanics perspectives, and continue to be analysed~\cite{versteegh2011gibbs,darrigol2018gibbs,allahverdyan2006explanation}. 
A crucial insight by Jaynes~\cite{jaynes1992gibbs} assuages our discomfort at the observer-dependent nature of the entropy change.
For an \emph{informed observer}, who sees the difference between the gases, the entropy increase has physical significance in terms of the work extractable through the mixing process -- in principle, they can build a device that couples to the two gases separately (for example, through a semi-permeable membrane) and thus let each gas do work on an external weight independently.
An \emph{ignorant observer}, who has no access to the distinguishing degree of freedom, has no device in their laboratory that can exploit the difference between the gases, and so cannot extract work.
For Jaynes, there is no paradox as long as one considers the abilities of the experimenter -- a viewpoint central to the present work.

A study of Gibbs mixing for identical quantum bosons or fermions is motivated by recognising that the laws of thermodynamics must be modified to account for quantum effects such as coherence~\cite{Lostaglio2015Description}, which can lead to enhanced performance of thermal machines~\cite{Vinjanampathy2016Quantum,Goold2016Role,Uzdin2015Equivalence}.
The thermodynamical implications of identical quantum particles have received renewed interest for applications such as Szilard engines~\cite{plesch2014maxwell,bengtsson2018quantum}, thermodynamical cycles~\cite{Myers2020Bosons,watanabe2020quantum} and energy transfer from boson bunching~\cite{holmes2020enhanced}.
Moreover, the particular quantum properties of identical particles, including entanglement, can be valuable resources in quantum information processing tasks \cite{killoran2014extracting,morris2019entanglement,Braun2018Quantum}

\begin{figure}
    \centering
    \includegraphics[width=0.4\textwidth]{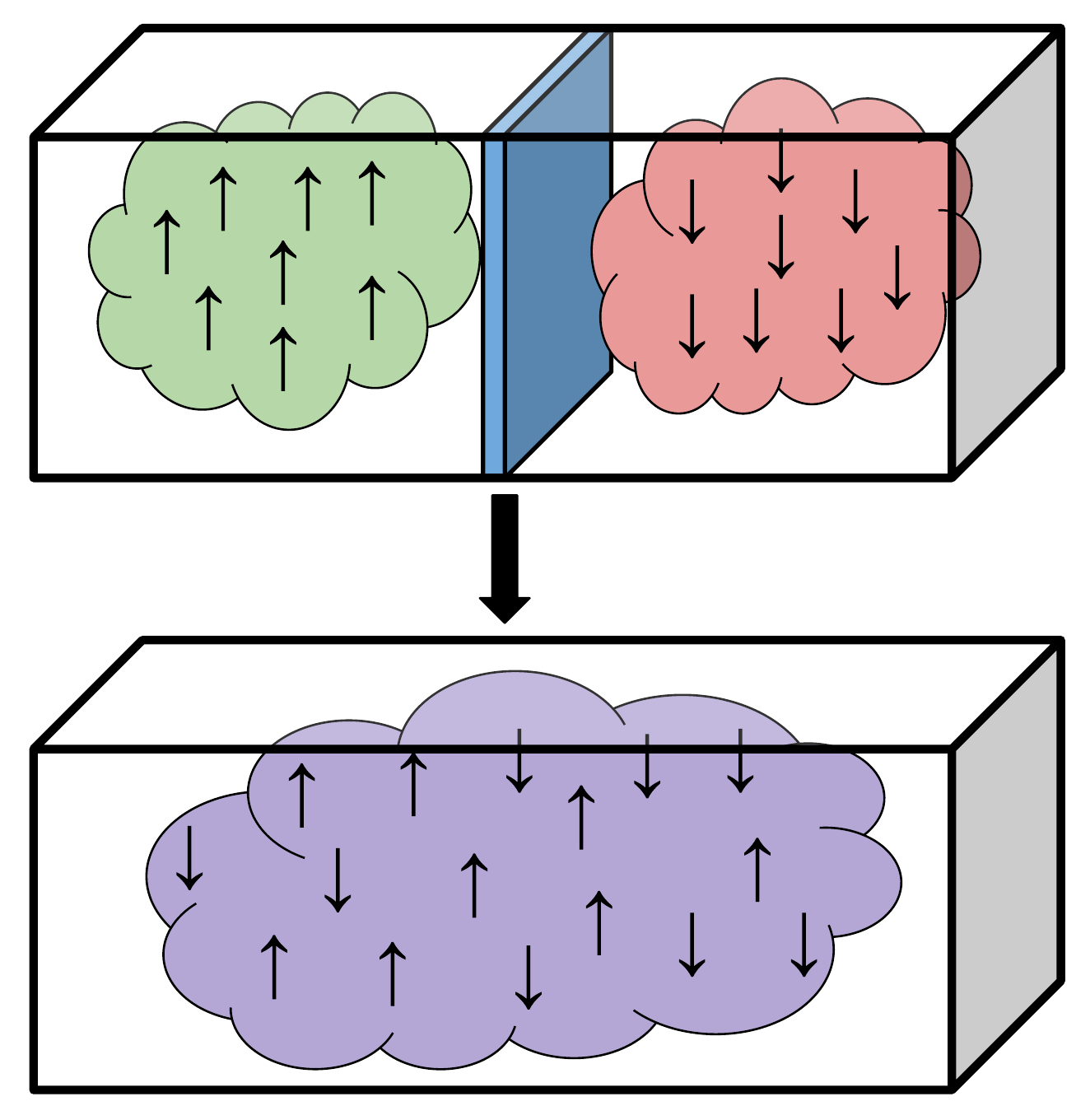}
    \caption{\textbf{The Gibbs paradox.} Two distinct gases of $n$ particles at the same temperature and pressure are separated by a partition.
    This partition is removed and the gases are allowed to mix and reach equilibrium.
    Two observers calculating the entropy increase during the process disagree depending on their ability to distinguish the particles.
    An informed observer, who can measure the difference between the gases, calculates $2n \ln 2$, while an observer ignorant of the difference records no entropy change.
    In this work, we ask how the situation changes when classical particles are replaced by identical quantum particles.
    }
    \label{fig:setting}
\end{figure}

In this work, we consider a toy model of an ideal gas with non-interacting quantum particles, distinguishing the two gases by a spin-like degree of freedom.
We describe the mixing processes that can be performed by both informed and ignorant observers, taking into account their different levels of control, from which we can calculate the corresponding entropy changes and thus work extractable by each observer.
For the informed observer, we recover the same results as obtained by classical statistical mechanics arguments.
However, for the ignorant observer, there is a marked divergence from the classical case.
Counter-intuitively, the ignorant observer can typically extract more work from distinguishable gases -- even though they appear indistinguishable -- than from truly identical gases.
In the continuum and large particle number limit which classically recovers the ideal gas, this divergence is maximal: \emph{the ignorant observer can extract as much work from apparently indistinguishable gases as the informed observer}.
Our analysis hinges on the symmetry properties of quantum states under permutations of particles.
For the ignorant observer, these properties lead to non-trivial restrictions on the possible work extraction processes.
Viewed another way, the microstates of the system described by the ignorant observer are highly non-classical entangled states. 
This implies a fundamentally different way of counting microstates, and therefore computing entropies, from what is done classically or even in semi-classical treatments of quantum gases.
Therefore we uncover a genuinely quantum thermodynamical effect in the Gibbs mixing scenario.
\begin{tableBox}[label=tab:observers]{Summary of the observers' abilities}
    \centering
    \small
    \begingroup
    \setlength{\tabcolsep}{10pt}
    \renewcommand{\arraystretch}{1.7}
    \begin{tabular}{c " c | c} 
        \textbf{Observer}    &   \textbf{Can}    &    \textbf{Can't} \\ \thickhline 
        \bf{Informed}    &   Access the spin and spatial degrees of freedom        & Change the number of up or down spins \\ \hline
        \bf{Ignorant}    &   Access the spatial degree of freedom        &   Access the spin degree of freedom
    \end{tabular}\label{tab:observers}
    \endgroup
\end{tableBox}

\section*{Results} 
\inlineheading{Set-up}
We consider a gas of $N$ particles inside a box, such that each particle has a position degree of freedom, denoted $x$, and a second degree of freedom which distinguish{es} the gases.
Since we only consider the case of two types of gases, this is a two-dimensional degree of freedom and we refer to this as the ``spin" $s$ (although it need not be an actual angular momentum).
Classically, the two spin labels are $\uparrow,\downarrow$, and their quantum analogues are orthogonal states $\ket{\uparrow},\ket{\downarrow}$.

Following the traditional presentation of the Gibbs paradox, the protocol starts with two independent gases on different sides of a box: $n$ on the left and $m = N-n$ on the right (see Fig.~\ref{fig:setting}).
Each side is initially thermalised with an external heat bath $B$ at temperature $T$.

In our toy model, each side of the box consists of $d/2$ ``cells'' ($d$ is even) representing different states that can be occupied by each particle.
These states are degenerate in energy, such that the Hamiltonian of the particles vanishes.
This might seem like an unrealistic assumption; however, this model contains the purely combinatorial (or ``state-counting'') statistical effects, first analysed by Boltzmann~\cite{Boltzmann1877Uber}, that are known to recover the entropy changes for a classical ideal gas~\cite{darrigol2018gibbs,Saunders2018Gibbs,Dieks2018Gibbs} using {the principle of equal a priori probabilities.} 
One could instead think of this setting as approximating a non-zero Hamiltonian in the high-temperature limit, such that each cell is equally likely to be occupied in a thermal state. 
{Since the particle number is strictly fixed, we are working in the canonical ensemble (rather than the grand canonical ensemble).}

Work extraction can be modelled in various ways in quantum thermodynamics.
In the resource-theoretic  approach based on thermal operations~\cite{horodecki2013fundamental,brandao2013resource}, one keeps track of all resources by treating the system (here, the particles), heat bath and work reservoir (or battery) as interacting quantum systems.
The work reservoir is an additional system with non-degenerate Hamiltonian whose energy changes are associated with work done by or on the system (generalising the classical idea of a weight being lifted and lowered).

The gases on either side of the box start in a state of local equilibrium and via mixing approach global equilibrium.
We therefore consider the extractable work to be given by the difference in non-equilibrium free energy $F$ \cite{bergmann1955new} between initial and final states, where $F(\rho) = \langle E\rangle_\rho - k_B T S(\rho)$, $\langle E\rangle_\rho  = \tr(\rho H)$ being the mean energy (zero in our case) and $S(\rho) = -\tr(\rho \ln \rho)$ the von Neumann entropy in natural units.
The extractable work in a process that takes $\rho$ to $\rho'$ is then {
\begin{equation} \label{eqn:work_def}
	W \leq F(\rho) - F(\rho') = k_B T \left[ S(\rho') - S(\rho) \right].
\end{equation}
In a classical reversible process, the extractable work is equal to the change in free energies.
This is generally an over-simplification for small systems, in which work can be defined in various ways~\cite{niedenzu2019concepts} -- e.g. required to be deterministic in the resource theory context~\cite{horodecki2013fundamental} or as a fluctuating random variable~\cite{aaberg2013truly,Dahlsten2011Inadequacy}, requiring consideration of other varieties of free energy.
However, equation~\eqref{eqn:work_def} will turn out to be sufficient for our purposes in the sense of mean extractable work.
We find the inequality to be saturable using thermal operations and characterise fluctuations around the mean in the latter part of our results section.
} 

Our analysis compares the work extracted by two observers with different levels of knowledge: the informed observer, who can tell the difference between the two gases, and the ignorant observer, who cannot.
The difference between these observers is that the former has access to the spin degree of freedom $s$, whereas the latter does not (summarised in Table~\ref{tab:observers}).

It is important to point out that, for the informed observer, the spin acts as a ``passive'' degree of freedom, meaning that it can be measured but not actively changed.
In other words, the two types  of gases cannot be converted into each other.
This assumption is always implicitly present in discussions of the Gibbs paradox -- without it, the distinguishing degree of freedom would constitute another subsystem with its own entropy changes
\footnote{One could also describe the spin as an information-bearing degree of freedom \cite{landauer1961irreversibility}.
The question is whether the information encoded within the spin state has an impact upon the thermodynamics of mixing.}.
\\

\inlineheading{Classical case}
Classically, the microstates described by the informed observer are specified by counting how many particles exist with each position $x$ and spin $s$ -- since the particles are indistinguishable \cite{bach1996indistinguishable}.
The ignorant observer has a different state space given by coarse-graining these states -- the classical equivalent of ``tracing out'' the spin degree of freedom.
Thus the ignorant observer can extract only as much work from two different gases as from a single gas, recovering Jaynes' original statement \cite{jaynes1992gibbs}.
These intuitively obvious facts are shown by a formal construction of the state spaces in Supplementary Note~1.
Paralleling our later quantum treatment, this establishes that the classical and quantum cases can be compared fairly.

The amount of extractable work in the classical case can be straightforwardly argued by state counting.
Consider the gas initially on the left side -- the number of ways of distributing $n$ particles among $d/2$ cells is $\binom{n+d/2-1}{n}$.
In the thermal state, each configuration occurs with equal probability.
Therefore the initial entropy, also including the gas on the right, is $\ln \binom{n+d/2-1}{n} + \ln \binom{m+d/2-1}{m}$.
For distinguishable gases, each gas can deliver work independently, with an equal distribution over $\binom{n+d-1}{n}\binom{m+d-1}{m}$ configurations.
For indistinguishable gases, the final thermal state is described as an equal distribution over all ways of putting $N=n+m$ particles into $d$ cells, of which there are $\binom{N+d-1}{N}$.
Hence the entropy change in each case is
\begin{align}
	\Delta S  =  &\ln \binom{n+d-1}{n} + \ln \binom{m+d-1}{m} - \ln \binom{n+d/2-1}{n} \nonumber \\
		&  - \ln \binom{m+d/2-1}{m} \quad \text{(distinguishable),} \label{eqn:class_dist} \\
	\Delta S = & \ln \binom{N+d-1}{N} - \ln \binom{n+d/2-1}{n} \label{eqn:class_indist} \nonumber \\
		& - \ln \binom{m+d/2-1}{m} \quad \text{(indistinguishable).}
\end{align}
Note that $\Delta S \neq 0$ even in the indistinguishable case, which may seem at odds intuitively with the result for an ideal gas.
However, one can check that $\Delta S = O(\ln N)$ in the limit of large $d$ (whereby the box becomes a continuum) and large $N$.
This is negligible compared with the ideal gas expression of $N \ln 2$ for distinguishable gases \cite{fujita1991indistinguishability}  \footnote{See \cite[p.~43]{darrigol2018gibbs} for a more detailed discussion of this approximation.}.
(Due to a subtle technicality with classical identical particles, formulas~\eqref{eqn:class_dist},\eqref{eqn:class_indist} might be regarded as upper bounds to the true values -- see Supplementary Note~1.)
Note that a classical analogue of fermions can be made by importing the Pauli exclusion principle, so that two or more particles can never occupy the same cell.
This has the effect of replacing the binomial coefficients of the form $\binom{N+d-1}{N}$ in \eqref{eqn:class_dist} and \eqref{eqn:class_indist} by $\binom{d}{N}$.\\
\begin{figure*}
\centering
	\includegraphics[width=0.75\textwidth]{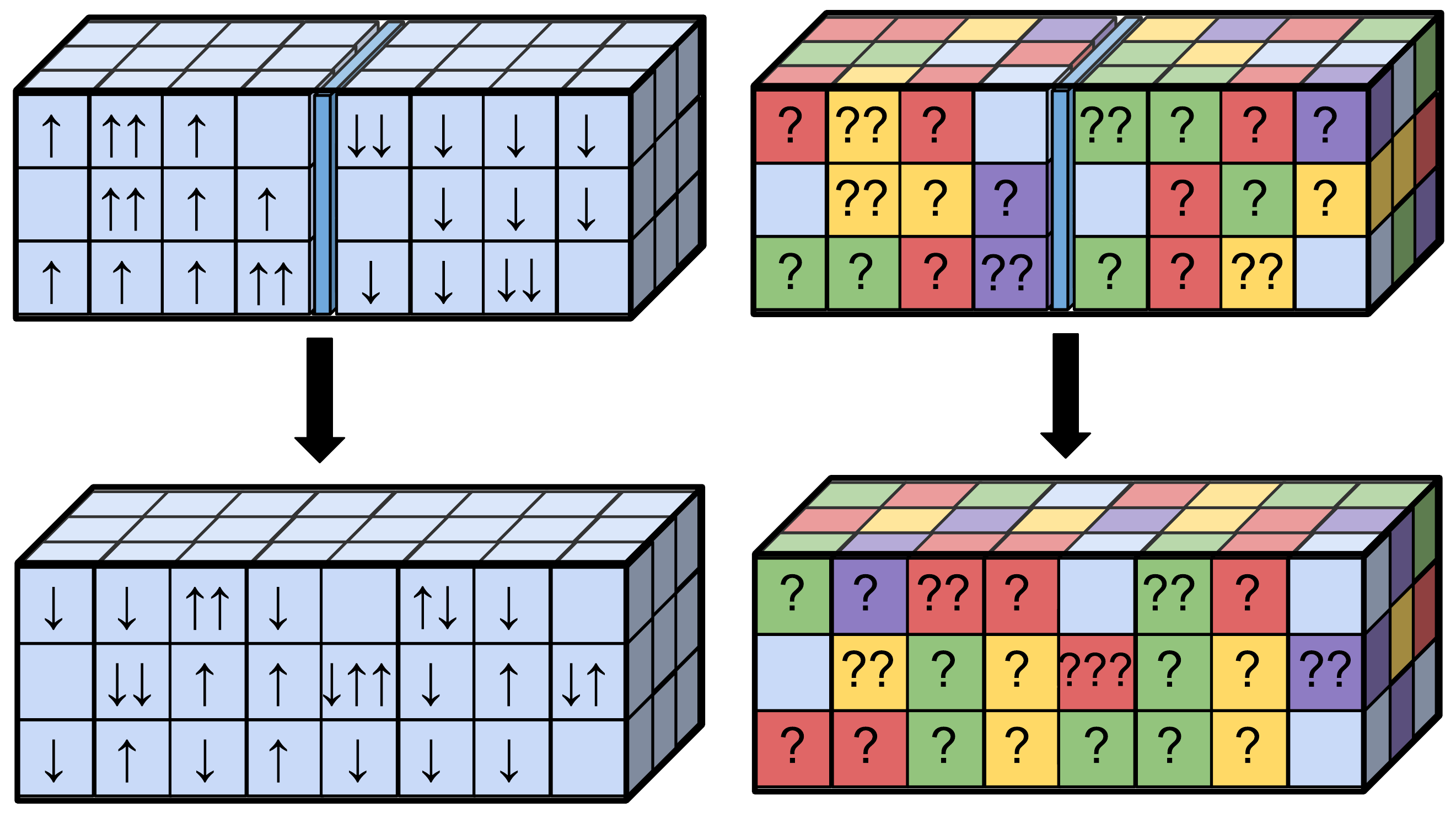}
		\caption{\textbf{Schematic of the quantum mixing process.} Two diagrams representing the mixing of indistinguishable (bosonic) quantum gases from the perspective of the informed (left) and ignorant (right) observers.
		Initially, $n$ spin-$\uparrow$ particles are found on the left and $m$ spin-$\downarrow$ on the right.
		The particles are then allowed to mix while coupling to an external heat bath and work reservoir.
		The informed observer describes microstates via the number of particles in each cell, and their respective spins.
		The ignorant observer cannot tell the spins states, but describes microstates (schematically depicted here by different colours) as superpositions of cell configurations, determined by the decomposition~\eqref{eqn:hilbert_space_decomp}. \label{fig:box1}}
\end{figure*}

\inlineheading{Quantum case}
Compared with the classical case, we must be more explicit about the role of the spin $s$ as a ``passive'' degree of freedom for the informed observer.
{This observer may obtain information about the numbers of spin-$\uparrow$ and spin-$\downarrow$ particles.
Thus they can engineer spin-dependent operations conditional on these numbers, but cannot change the number of each spin.
}

For identical gases, the result is of course the same as for the ignorant observer, and the classical case \eqref{eqn:class_indist}.
For distinguishable gases, each gas behaves as an independent subsystem; thus, the entropy changes are the same as for classical distinguishable gases \eqref{eqn:class_dist}.

The remainder of this section is devoted to the ignorant observer, for which we find a departure from the classical case.\\

The peculiarities of the quantum case stem from a careful look at the Hilbert space structure.
The Hilbert space of a single particle is a product $\mc{H}_1 = \mc{H}_x \ox \mc{H}_s$ of a part for the spatial degree of freedom $x$ and a part for the spin $s$.
Since there are $d$ cell modes and two spin states, these parts have dimensions $\dim \mc{H}_x=d,\, \dim \mc{H}_s=2$.
For $N$ distinguishable particles, the state space would be $\mc{H}_1^{\ox N}$.
However, for bosons and fermions, which are quantum indistinguishable particles, states lie in the symmetric and antisymmetric subspaces, respectively (in first quantisation).
This symmetry refers to the wavefunction under permutations of particles: for bosons, there is no change, whereas for fermions, each swap of a pair incurs a minus sign in the global phase.
The physical Hilbert space of $N$ particles can then be written as
\begin{align}
	\mc{H}_N = P_{\pm} \left( \mc{H}_x^{\ox N} \ox \mc{H}_s^{\ox N} \right),
\end{align}
where $P_{+(-)}$ is the projector onto the (anti-)symmetric subspace.

Since each particle carries a position and spin state, a permutation $\Pi$ of particles is applied simultaneously to these two parts: $\Pi$ acts on the above Hilbert space in the form $\Pi_x \ox \Pi_s$.
The requirement of an overall (anti-)symmetric wavefunction effectively couples these two degrees of freedom via their symmetries.
For a familiar example, consider two particles.
The spin state space can be broken down into the symmetric ``triplet'' subspace spanned by $\ket{\uparrow\uparrow},\, \ket{\downarrow\downarrow}$ and $\ket{\uparrow\downarrow}+\ket{\downarrow\uparrow}$, and the antisymmetric ``singlet'' subspace consisting of $\ket{\uparrow\downarrow}-\ket{\downarrow\uparrow}$.
For bosons, overall symmetry requires that a triplet spin state be paired with a symmetric spatial wavefunction, and a singlet spin state with an antisymmetric spatial function.
For fermions, opposite symmetries are paired.

With more particles, the description is more complex, but the main idea of paired symmetries remains the same.
Following~\cite{adamson2008detecting}, our main tool is Schur-Weyl duality~\cite{GoodmanWallach}, which decomposes
\begin{equation} \label{eqn:schur-weyl}
	\mc{H}_x^{\ox N} = \bigoplus_\lambda \mc{H}_x^\lambda \ox \mc{K}_x^\lambda,
\end{equation}
where $\lambda$ runs over all Young diagrams of $N$ boxes and no more than $d$ rows \footnote{A Young diagram can be described simply by a non-increasing set of ($\leq d$) positive integers summing up to $N$}.
In technical terms, $\mc{H}_x^\lambda$ and $\mc{K}_x^\lambda$ carry irreducible representations of the unitary group $\mathrm{U}(d)$ and the permutation group $S_N$ of $N$ particles, respectively.
More concretely, a non-interacting unitary operation on the positions of all the particles, $u_x^{\ox N}$, is represented in the decomposition~\eqref{eqn:schur-weyl} as an independent rotation within each of the $\mc{H}_x^\lambda$ spaces.
The term ``irreducible'' refers to the fact that each of these spaces may be fully explored by varying the unitary $u_x$.
Similarly, a permutation of the particles in the spatial part of the wavefunction is represented by an action on each $\mc{K}_x^\lambda$ space.
Thus each block labelled by $\lambda$ in the decomposition~\eqref{eqn:schur-weyl} has a specific type of permutation symmetry.

The same decomposition works for the spin part $\mc{H}_s^{\ox N}$.
However, since this degree of freedom is two-dimensional, each $\lambda$ is constrained to have no more than two rows.
We can think of $s$ as describing a total angular momentum formed of $N$ spin-$1/2$ particles, and in fact $\lambda$ can be replaced by a total angular momentum eigenvalue $J$ varying over the range $N/2, N/2-1, \dots$.

After putting the spatial and spin decompositions together, projecting onto the overall (anti-)symmetric subspace causes the symmetries of the two parts to be linked.
For bosons, the $\lambda$ label for $x$ and $s$ must be the same; for fermions, they are transposes of each other (i.e, related by interchanging rows and columns).
This results in the form
\begin{align} \label{eqn:hilbert_space_decomp}
	\mc{H}_N & = \bigoplus_\lambda \mc{H}_x^\lambda \ox \mc{H}_s^\lambda \quad \text{for bosons,} \nonumber \\
	\mc{H}_N & = \bigoplus_\lambda \mc{H}_x^{\lambda^T} \ox \mc{H}_s^\lambda \quad \text{for fermions.}
\end{align}
Instead of the label $\lambda$, from now on we use the angular momentum number $J$ and generally write this decomposition as $\bigoplus_J \mc{H}_x^J \ox \mc{H}_s^J$ -- bearing in mind that $\mc{H}_x^J$ is different for bosons and fermions.
In terms of the earlier $N=2$ example, $J=1$ corresponds to the spin triplet subspace, and $J=0$ to the spin singlet.

Another way of describing the decomposition~\eqref{eqn:hilbert_space_decomp} is that it provides a convenient basis ${\ket{J,q}}_x {\ket{J,M}}_s {\ket{\phi_J}}_{xs}$, known as the Schur basis \cite{harrow2005applications}.
Here, $\{{\ket{J,q}}_x\}_q$ is a basis for $\mc{H}_x^J$ and $\{{\ket{J,M}}_s\}_M$ a basis for $\mc{H}_s^J$.
$M=-J,-J+1,\dots,J$ can be interpreted as the total angular momentum quantum number along the $z$-axis.
${\ket{\phi_J}}_{xs} \in \mc{K}_x^J \ox \mc{K}_s^J$ is a state shared between the $x$ and $s$ degrees of freedom.\\

We now consider how the state thermalises for the ignorant observer.
Since the ignorant observer cannot interact with spin, their effective state space is described by tracing out the factor $\mc{H}_s$ for each particle.
In terms of the decomposition~\eqref{eqn:hilbert_space_decomp} and corresponding basis described above, this means that an initial density matrix $\rho$, after tracing out $s$, is of the form
\begin{equation} \label{eqn:initial_state}
	\rho_x := \tr_s \rho = \bigoplus_J p_J \rho_x^J \ox \tr_s {\proj{\phi_J}}_{xs},
\end{equation}
where $\rho_x^J$ is a density matrix on $\mc{H}_x^J$, occurring with probability $p_J$.
Note that there is no coherence between different values of $J$, and that the components $\rho_x^J$ are mutually perfectly distinguishable by a measurement of their $J$.

Additionally, the allowed operations must preserve the bosonic or fermionic exchange symmetry.
Any global unitary $U_{xBW}$, coupling the spatial degree of freedom of the particles to the heat bath and work reservoir, must therefore commute with permutations on the spatial part: $[U_{xBW}, \Pi_x] = 0$ for all $\Pi$.
By Schur's Lemma, such a unitary decomposes as $U = \bigoplus_J U^J \ox I^J$, where $U^J$ operates on the $\mc{H}_x^J$ component, with an identity $I^J$ on $\mc{K}_x^J$.
Hence each $J$ component is operated upon independently, the spin eigenvalue $J$ being conserved.

In summary, therefore, the ignorant observer may engineer any thermal operation extracting work separately from each $J$ component (depicted in Fig.~\ref{fig:box1}).
We can think of their operations being conditioned on the spatial symmetry type, and although $J$ is observed to fluctuate randomly, a certain amount of work is extracted for each $J$ (see the latter part of the results section for a more detailed analysis of this fluctuation). For each $J$, there exists a operation within the thermal operations framework~\cite{horodecki2013fundamental} that performs deterministic work extraction saturating inequality~\eqref{eqn:work_def}.
This is because the transformation is between (energy-degenerate) uniformly mixed states of differing dimension \footnote{Note that the work extraction process does not involve a measurement by the observer -- only a coupling to the apparatus that depends on the value of $J$. Therefore there is no need to consider an additional entropic measurement cost, unlike the case of Maxwell's demon~\cite{zurek1986maxwell,bennett2003notes}}.

The question of optimal work extraction thus reduces to calculating the entropy of the initial state~\eqref{eqn:initial_state} and finding the maximum entropy final state.
The fully thermalised final state seen by the ignorant observer is maximally mixed within each $J$ block:
\begin{equation} \label{eqn:final_state}
	\rho'_x = \bigoplus_J p_J \frac{I^J_x}{d_J} \ox \tr_s {\proj{\phi_J}}_{xs},
\end{equation}
where $I^J_x$ is the identity on $\mc{H}_x^J$ and $d_J$ is the corresponding dimension.

The overall entropy change is the average over all $J$, found to be (with details in Supplementary Note~2):
\begin{align} \label{eqn:entropy_bosons}
	\DS{igno} &= \sum_J p_J \DS{igno}^J, \nonumber \\
			& = \sum_J p_J \ln d^B_J - \ln \binom{n+d/2-1}{n} - \ln \binom{m+d/2-1}{m}
\end{align}
for bosons, and
\begin{align} \label{eqn:entropy_fermions}
	\DS{igno} &=  \sum_J p_J \ln d^F_J - \ln \binom{d/2}{n} - \ln \binom{d/2}{m}
\end{align}
for fermions.
Expressions for the dimensions $d^{B,F}_J$ are found in Supplementary Note~4:
\begin{align} \label{eqn:dimensions}
	d^B_J & = \frac{(2J+1) \left(\frac{N}{2}-J+d-2\right)! \left( \frac{N}{2}+J+d-1 \right)!}{\left(\frac{N}{2}-J\right)! \left(\frac{N}{2}+J+1\right)! (d-1)!(d-2)!}, \nonumber \\
	d^F_J & = \frac{(2J+1) d! (d+1)! }{\left(\frac{N}{2}+J+1\right)! \left(\frac{N}{2}-J\right)! \left(d-\frac{N}{2}+J+1\right)! \left(d-\frac{N}{2}-J\right)! }.
\end{align}
The probabilities $p_J$ are found (see Supplementary Note~2) from the Clebsch-Gordan coefficients $C(j_1,m_1;j_2,m_2;J,M)$ describing the coupling of two spins with angular momentum quantum numbers $(j_1,m_1),\, (j_2,m_2)$ into overall quantum numbers $(J,M)$.
Here, the two spins are the groups of particles on the left and right, respectively.

For identical gases, all particles have spins in the same direction, so the spin wavefunction is simply ${\ket{\uparrow}}^{\ox N}$.
This state lies fully in the subspace of maximal total spin eigenvalue, $J=M=N/2$ -- which is also fully symmetric with respect to permutations.
Thus the spin part factorises out (i.e., there is no correlation between spin and spatial degrees of freedom).
It is then clear that dimension counting reduces to the classical logic of counting ways to distribute particles between cells.
Indeed, the dimension of the subspace $\mc{H}_x^{N/2}$ is $d^B_{N/2} = \binom{N+d-1}{N}$ for bosons and $d^F_{N/2} = \binom{d}{N}$ for fermions.
It follows that we recover the entropy as the classical case of indistinguishable particles~\eqref{eqn:class_indist}.

For orthogonal spins, there are $n$ spin-$\uparrow$ and $m$ spin-$\downarrow$, leading to $M=(n-m)/2$ and a distribution over different values of $J$ according to
\begin{align} \label{eqn:pj_cg_coeff}
	p_J & = \frac{(2J+1) n! m!}{\left(\frac{N}{2}+J+1\right)! \left(\frac{N}{2}-J\right)!}.
\end{align}
The resulting entropies and significant limits are discussed after an example. \\

\begin{tableBox}[label=tab:results]{Summary of results}
    {\centering
    \small
    \begingroup
    \setlength{\tabcolsep}{3pt}
    \renewcommand{\arraystretch}{1.8}
    \begin{tabular}{c " c | c | c | c | c} 
        & \textbf{Quantum}		& \textbf{Classical}		& \textbf{Quantum}	& \textbf{Quantum}& \textbf{Classical}			 	\\ 
				\textbf{Limit} & (no limit)		& (no limit)		& $(d\gg n^2)$		& $(d\gg n^2\gg 1)$ &  $(d\gg n^2\gg 1)$			 	\\ \thickhline 
			$\DS{info}$	& $2 \ln \binom{n+d-1}{n}-2\ln \binom{n+d/2-1}{n}$	& $2\ln\binom{n+d-1}{n}-2\ln\binom{n+d/2-1}{n}$	& $\dots$	& $\approx 2n\ln 2$  & $\approx 2n\ln 2$	\\ \hline
        $\DS{igno}$		&  $\sum_J p_J \ln d^{B}_J - 2\ln \binom{n+d/2-1}{n}$	& $\ln\binom{2n+d-1}{2n}-2\ln\binom{n+d/2-1}{n}$ 	& $\approx \DS{info} - H(\bg{p}) - \frac{n^2}{2d^2}$	& $\approx 2n\ln 2 $ & $\approx 0$
    \end{tabular} \vspace{10pt}
    \endgroup}

    Entropy changes $\DS{info},\DS{igno}$ for the informed and ignorant observers and their limits are expressed for bosons with $n=m$. For fermions, replace the dimension of the symmetric subspace $\binom{n+d-1}{n}$ with that of the antisymmetric one $\binom{d}{n}$ and $d^{B}_J$ by  $d^{F}_J$ (both of which are defined in equation~\eqref{eqn:dimensions}).
\end{tableBox}

\inlineheading{Example}
Taking $n=m=1$ demonstrates the mechanism behind the state space decomposition.
For two particles, there are only two values of $J$, corresponding to the familiar singlet and triplet subspaces:
\begin{align}
	\mc{H}_s^0 & = \mathrm{span} \left\{ \ket{\uparrow\downarrow}-\ket{\downarrow\uparrow}  \right\}, \nonumber \\
	\mc{H}_s^1 & = \mathrm{span} \left\{  \ket{\uparrow \uparrow}, \ket{\downarrow \downarrow}, \ket{\uparrow\downarrow} + \ket{\downarrow\uparrow} \right\}.
\end{align}
Consider a spatial configuration where a spin-$\uparrow$ particle is on the left in cell $i$, and a spin-$\downarrow$ is on the right in cell $j$.
For bosons, the properly symmetrised wavefunction is
\begin{align}
	\ket{\psi_{i,j}} & := \frac{1}{\sqrt{2}}\left( {\ket{i_L j_R}}_x{\ket{\uparrow\downarrow}}_s + {\ket{j_R i_L}}_x{\ket{\downarrow\uparrow}}_s \right) \nonumber \\
		& = \frac{1}{\sqrt{2}} \left[ \frac{\ket{i_L j_R}-\ket{j_R i_L}}{\sqrt{2}} \cdot \frac{\ket{\uparrow\downarrow} -\ket{\downarrow\uparrow}}{\sqrt{2}} \right. \quad (J=0) \nonumber \\
		& \qquad \left. + \frac{\ket{i_L j_R}+\ket{j_R i_L}}{\sqrt{2}} \cdot \frac{\ket{\uparrow\downarrow}+\ket{\downarrow\uparrow}}{\sqrt{2}} \quad (J=1)\right].
\end{align}
So $p_0=p_1=1/2$, and the spatial component of this state is conditionally pure for both $J$.
The initial thermal state is a uniform mixture of all such $\ket{\psi_{i,j}}$, with $(d/2)^2$ terms.
Thus $S(\rho_x^0) = S(\rho_x^1) = 2(\ln d-\ln2)$.
For the final thermal state, we observe that
\begin{align}
	\mc{H}_x^0 & = \mathrm{span} \left\{ \ket{ij}-\ket{ji} \mid i<j \right\}, \nonumber \\
	\mc{H}_x^1 & = \mathrm{span} \left\{ \ket{ij}+\ket{ji} \mid i\leq j \right\},
\end{align}
where $i,j$ now label cells either on the left or right.
The corresponding dimensions are $d_0 = d(d-1)/2,\, d_1 = d(d+1)/2$.
Within the $J=0$ subspace, the entropy change is $\ln[d(d-1)/2]-2\ln d + 2\ln 2 = \ln(1-1/d) + \ln 2$, and for $J=1$, it is $\ln[d(d+1)]-2\ln d +2\ln 2= \ln(1+1/d)+\ln 2$.
Overall, therefore,
\begin{align}
	\Delta S_\text{igno} & = \frac{1}{2}\ln\left(1-\frac{1}{d}\right) + \frac{1}{2}\ln \left(1+\frac{1}{d}\right) + \ln 2 \nonumber \\
		& = \frac{1}{2} \ln\left(1-\frac{1}{d^2}\right) +  \ln 2.
\end{align}
For the informed observer, we have $\Delta S_\text{info} = 2\ln 2$.
For identical gases, we find $\Delta S_\text{iden} = \ln(1+1/d)+ \ln 2$, strictly greater than $\Delta S_\text{igno}$, but the two become equal in the limit $d \to \infty$.

Repeating the same calculation with fermions, the symmetric and antisymmetric states now pair up oppositely. Then $\Delta S_\text{igno}$ is the same as for bosons.
However, we have $\Delta S_\text{iden} = \ln(1 - 1/d) + \ln 2 < \Delta S_\text{igno}$. Unlike for bosons, two distinguishable fermions permit more extractable work by the ignorant observer than two identical fermions.\\

\begin{figure*}[t]
\centering
         \includegraphics[width=\textwidth]{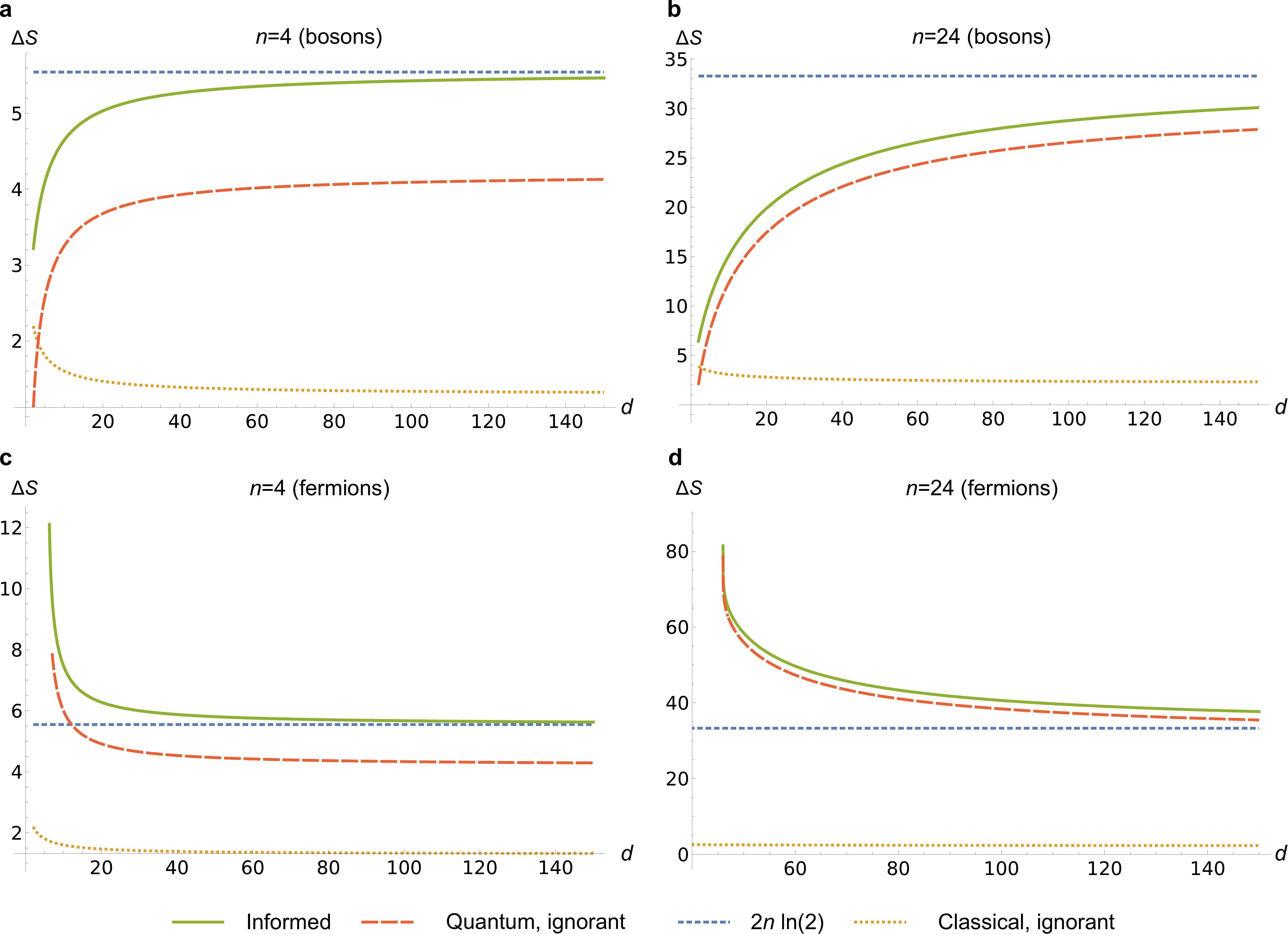} 
                \caption{\textbf{Entropy changes as a function of dimension.} Series of plots showing $\DS{info},\DS{igno}$ against the total cell number $d$ of the system. \textbf{a, b)} bosonic systems of particle number $n=4$ and $n=24$ respectively. \textbf{c, d)} the same for fermionic systems. Note that we have taken the initial number of particles on either side of the box to be equal, $n=m$ in all cases. For comparison, all four figures also display the classical changes in entropy for an informed/ignorant observer. The behaviour of the deficit between $\Delta S$ for an informed/ignorant observer of quantum particles agrees with the low density limit in equation \eqref{eqn:boson_low_density} where  we can see $\DS{info}$ tending to the classical limit $2n \ln (2)$ with $\DS{igno}$ trailing behind by a deficit of $n^2/d^2+H(\bg{p})$. Additionally, by comparing the different plots, we can see the low-dimensional fermionic advantage where the change in entropy is even greater than the classical $2n \ln (2)$ value. \label{fig:four_figures} }
        \label{fig:four_figures}
\end{figure*}

\inlineheading{Entropy changes and limits}\label{sec:results}
In Fig.~\ref{fig:four_figures} we plot both  $\DS{info}$ and $\DS{igno}$ as a function of dimension for bosons and fermions.
Below we analyse the special cases and limits which emerge from these expressions, summarised in Table~\ref{tab:results}.\\

With bosons, there are two special cases in which it is easily proven that distinguishable gases are less useful than indistinguishable ones for the ignorant observer.
The first case is the example above, with $n=m=1$.
In addition, for $d=2$, we have $d^B_J = 2J+1$ -- so the largest subspace is that with maximal $J = N/2$.
The largest entropy change is then obtained when $p_{N/2} = 1$, which is satisfied precisely for indistinguishable gases.

For fermions, we see from Fig.~\ref{fig:four_figures} that the greatest work -- for both observers -- is obtained for small $d$.
An intuitive explanation is that the Pauli exclusion principle causes the initial state to be constrained and thus have low entropy.
For example, with the minimal dimension $d=2n=2m$, we have $\DS{info} = 2\ln \binom{2n}{n} \approx 4n \ln 2$ to leading order when $n$ is large.
The ignorant observer can do almost as well: the state is entirely contained in the $J=0$ subspace, with
$
	d^F_0 = \frac{(2n)!(2n+1)!}{(n!)^2 (n+1)!^2} = \frac{2n+1}{(n+1)^2} \binom{2n}{n}^2,
$
giving $\DS{igno} \approx 4n \ln 2$ for large $n$.
This is twice as much as for the classical ideal gas.\\

The most interesting conclusion is reached in the limit of large $d \gg n^2$, which we term the \emph{low density limit}.
For simplicity, we take $n=m$.
To lowest order in $n^2/d$, we find
\begin{equation} \label{eqn:boson_low_density}
	\DS{igno} \approx \DS{info} - H(\bg{p}) - \frac{n^2}{2d^2},
\end{equation}
where $H(\bg{p}) = -\sum_J p_J \ln p_J$ is the Shannon entropy of the distribution $p_J$.
Thus, as $d \to \infty$, the ignorant observer can extract as much work as the informed one, minus an amount $H(\bg{p})$.
This gap is evident from the graphs in Fig.~\ref{fig:four_figures}.

Now consider the limit $d \gg n^2, \, n \gg 1$, with both low density and large particle number.
Classically, this limit recovers ideal gas behaviour -- the large dimension limit can be thought of as letting the box become a continuum.
In Supplementary Note~6, we show that $H(\bg{p})$ (which depends only on $n$, not $d$), behaves as
\begin{equation}
	H(\bg{p}) \approx \frac{1}{2} \ln n + 0.595...,
\end{equation}
with a correction going to zero as $n \to \infty$.
Recall that the entropy change for the informed observer is approximately $2n \ln 2$ in this limit.
Therefore the deficit $H(\bg{p})$, which is logarithmic, becomes negligible compared with $2n \ln 2$.
Thus the ignorant observer can extract essentially as much work as the informed observer: $\DS{igno} \approx \DS{info} \approx 2n \ln 2$.
This result is remarkable because it shows a macroscopic departure from the classical case in this limit.\\

How can we understand this low density limit? An important feature of the low density limit is that the final entropy becomes as large as it could possibly be: $\rho'_x$ becomes maximally mixed over its whole state space.
This is true for any $N$, not just large numbers.
We now give an explanation of this phenomenon, which proceeds by counting the number of mutually orthogonal states which can be accessed by the ignorant observer.

The important point about the low density limit is that particles almost never sit on top of each other -- that is, almost all states are such that precisely $N$ cells are occupied, each with a single particle.
More formally, the number of ways of putting $N$ bosonic particles into $d$ cells is $\binom{N+d-1}{N} \approx \binom{d}{N}$ when $d$ is large, where the approximation means the ratio of the two sides is close to unity.
Let us refer to each of these $\binom{d}{N}$ choices of (singly) occupied cells as a \emph{cell configuration}.
For each cell configuration, there are $\binom{N}{n}$ \emph{spin configurations}, i.e., ways of distributing the $n$ spin-$\uparrow$ and $m$ spin-$\downarrow$ particles.
In classical physics, the ignorant observer cannot distinguish any of the spin configurations corresponding to a single cell configuration.
In quantum mechanics, remarkably, there are precisely $\binom{N}{n}$ states which can be fully distinguished by the ignorant observer, each being a superposition of different spin configurations.

Let us choose a single cell configuration -- without loss of generality, let cells $1,\dots,N$ be occupied.
The state of a spin configuration is denoted as a permutation of
\begin{equation}
	{\ket{\uparrow}}_1 \dots {\ket{\uparrow}}_n {\ket{\downarrow}}_{n+1} \dots {\ket{\downarrow}}_N \in (\mathbb{C}^2)^{\ox N},
\end{equation}
where each cell is treated as a qubit with basis states $\ket{\uparrow},\ket{\downarrow}$ according to which type of spin occupies it.
(Note that the subsystems being labelled are here are the occupied cells, not particles.)

Again using Schur-Weyl duality, the state space of $N$ qubits can be decomposed as
\begin{equation}
	(\mathbb{C}^2)^{\ox N} = \bigoplus_J \mc{H}^J \ox \mc{K}^J.
\end{equation}
Due to this decomposition, there is a natural basis $\ket{J, M, p}$, where $\mathrm{SU}(2)$ spin rotations $u_s^{\ox N}$ act on the $M$ label (denoting the eigenvalue of the total $z$-direction spin), and permutations $\Pi$ of the $N$ cells act on the $p$ label.

How do we represent the effective state seen by the ignorant observer?
In the representation used here, this corresponds to \emph{twirling} over the spin states, i.e., performing a Haar measure average over all spin rotations $u_s^{\ox N}$~\cite{bartlett2007reference}.
In the basis $\ket{J,M,p}$, however, this is a straightforward matter of tracing out the $\mc{H}^J$ subspaces, since only these are acted on by the twirling operation.
Thus the ignorant observer has access to states labelled as $\ket{J,p}$.

How much information has been lost by tracing out $\mc{H}^J$?
In fact, none -- the label $M = (n-m)/2$ is fixed.
Therefore the experimenter can perfectly distinguish all the basis states $\ket{J,p}$ -- and there are just as many of these as there are spin configurations, namely $\binom{N}{n}$.

For example, take $n=m=1$: the two spin configurations are $\ket{\uparrow\downarrow}, \ket{\downarrow\uparrow}$, and for some pair of occupied cells, the two distinguishable states are
\begin{align}
	\ket{J=1,\, M=0,\, p=0} & = \frac{1}{\sqrt{2}} \left( \ket{\uparrow\downarrow} + \ket{\downarrow\uparrow} \right), \nonumber \\
	\ket{J=0,\, M=0,\, p=0} & = \frac{1}{\sqrt{2}} \left( \ket{\uparrow\downarrow} - \ket{\downarrow\uparrow} \right).
\end{align}
Since these are respectively in the triplet and singlet subspaces, they remain orthogonal even after twirling.
They can be distinguished by mixing the cells at a balanced beam splitter: it is easy to show that the symmetric state ends up with a superposition of both particles in cell 1 and both in cell 2, while the antisymmetric state ends up with one particle on each side.
Therefore, after this beam splitter, the two states can be distinguished by counting the total particle number in each cell.

A slightly more complex example is with $n=2,m=1$.
Then the distinguishable basis states for three occupied cells are
\begin{align}
	\ket{J=\frac{3}{2},\, M=\frac{1}{2},\, p=0} & = \frac{1}{\sqrt{3}} \left( \ket{\uparrow\uparrow\downarrow} + \ket{\uparrow\downarrow\uparrow} + \ket{\downarrow\uparrow\uparrow} \right), \nonumber \\
	\ket{J=\frac{1}{2},\, M=\frac{1}{2},\, p=0} & = \frac{1}{\sqrt{2}} \left( \ket{\downarrow\uparrow\uparrow} + \ket{\uparrow\downarrow\uparrow} \right),   \\
	\ket{J=\frac{1}{2},\, M=\frac{1}{2},\, p=1} & = \sqrt{\frac{2}{3}} \ket{\uparrow\uparrow\downarrow} - \frac{1}{\sqrt{6}} \left( \ket{\uparrow\downarrow\uparrow} + \ket{\downarrow\uparrow\uparrow} \right). \nonumber
\end{align}

Observe that the argument in this section does not depend in anyway on the exchange statistics of the particles, explaining why we see the same limit for bosons and fermions.\\

\inlineheading{Quantumness of the protocol}
The above discussion of the low density limit clarifies the fundamental reason why the quantum ignorant observer performs better than the classical one.
The distinguishable states comprising the final thermalised state are superpositions of different spin configurations.
We might describe a classical observer within the quantum setting as one who is limited to operations diagonal in the basis of cell configurations -- that is, they are only able to count the number of particles occupying each cell.
For such an observer, these superposition states are indistinguishable.

A crucial question is then: how difficult is it to engineer the quantum protocol for the ignorant observer?
We can imagine that the heat bath and work reservoir might naturally couple to the system in the cell occupation basis (if this is the basis that emerges in the classical case).
The required coupling is in the Schur basis ${\ket{J,q}}_x$, which are generally highly entangled between cells.
A sense of their complexity is given by the unitary that rotates the Schur basis to the computational basis, known as the Schur transform.
Efficient algorithms to implement this transform have been found~\cite{bacon2006efficient}, with a quantum circuit whose size is polynomial in $N, d, \ln(1/\epsilon)$, allowing for error $\epsilon$.
This circuit is related to the quantum Fourier transform, an important subroutine in many quantum algorithms.
Thus, while the Schur transform can be implemented efficiently, it appears that engineering the required work extraction protocol -- in the absence of fortuitous symmetries in the physical systems being used -- may be as complex as universal quantum computation.\\

\inlineheading{Work fluctuations}\label{sec:fluctuations}
The work extraction protocol we have presented is not deterministic: for each value of $J$, a different amount of work is extracted with probability $p_J$.
This is typically expected of thermodynamics of small systems; however, in classical macroscopic thermodynamics, such fluctuations are negligible.
We can ask whether the same is true of the work extracted by the ignorant observer in the quantum case, especially in the low density and large particle number limits.

One informative way of quantifying the fluctuations is via the variance of entropy change.
Let us denote the entropy change for each $J$ by $\DS{igno}(J)$.
The mean is $\DS{igno} = \sum_J p_J \DS{igno}(J)$, and the variance is $V(\DS{igno}) = \sum_J p_J \DS{igno}(J)^2 - \DS{igno}^2$.
This can be computed straightforwardly from our expressions for $p_J,d_J$, and approximated in various limits.

Consider first a high density BEC-limit case with $d=2$ and $N=2n \gg 1$ bosons.
We have $d^B_J = 2J+1$, and using the techniques of Supplementary Note~6, $p_J \approx \frac{2J}{n} e^{-J^2/n}$.
Then $\DS{igno} = \sum_J p_J \ln(2J+1) \approx \frac{1}{2} \ln n + \ln 2 - \frac{\gamma}{2} \approx \frac{1}{2} \ln n + 0.405$.
Similarly, we compute $V(\DS{igno}) = \sum_J p_J [\ln(2J+1)]^2 \approx \frac{\pi^2}{24} \approx 0.411$.
Therefore the mean work dominates its fluctuations (logarithmic versus a constant).

Next, consider the closest analogue for fermions: the case of minimal dimension $d=2n=2m$.
Recall that $\DS{igno} \approx \DS{info} \approx 4n \ln 2$ for large $n$.
Since $p_0=1$, work extraction is in fact completely deterministic in this case.

Finally, take the low density limit.
As found before, for both bosons and fermions, $\DS{igno} \approx 2n \ln 2$ -- linear in $n$ -- and yet we still find a constant $V(\DS{igno}) \approx \frac{\pi^2}{24}$.

In these macroscopic limits, therefore, work extraction is either fully deterministic or effectively deterministic in that the fluctuations are negligible compared with the mean.\\

\inlineheading{Non-orthogonal spins}
The results generalise to the case of partially distinguishable spins -- that is, initially with $n$ in spin state $\ket{\uparrow}$ on the left and $m$ in state $\ket{\nearrow}$ on the right, where
\begin{equation} \label{eqn:rotated_spin}
	\ket{\nearrow} = \cos(\theta/2)\ket{\uparrow} + \sin(\theta/2) \ket{\downarrow}.
\end{equation}
For this, we must be more explicit about the operations permitted by the informed observer.
The most general global unitary that does not affect the number of each type of spin is of the form $U = \bigoplus_M U^{(M)}_{xsBW}$, where the block structure refers to subspaces with fixed $M$ as defined by the Schur basis (recalling that the total number of particles is fixed).
We find (see Supplementary Note~3 for details) that $\DS{info}$ is an average of entropy changes for each value of $M$.
For $\DS{igno}$, all that changes is the probability $p_J$, now being obtained by an average over Clebsch-Gordan coefficients.
Importantly, for both observers, the result is a function of $\theta$ only via the probability distribution $q_M$ for the spin value $M$.
In Fig.~\ref{fig:partial_plot}, one observes the smooth transition from identical to orthogonal spin states as $\theta$ varies from $0$ to $\pi$.

\begin{figure}[h!]
        \includegraphics[width=0.43\textwidth]{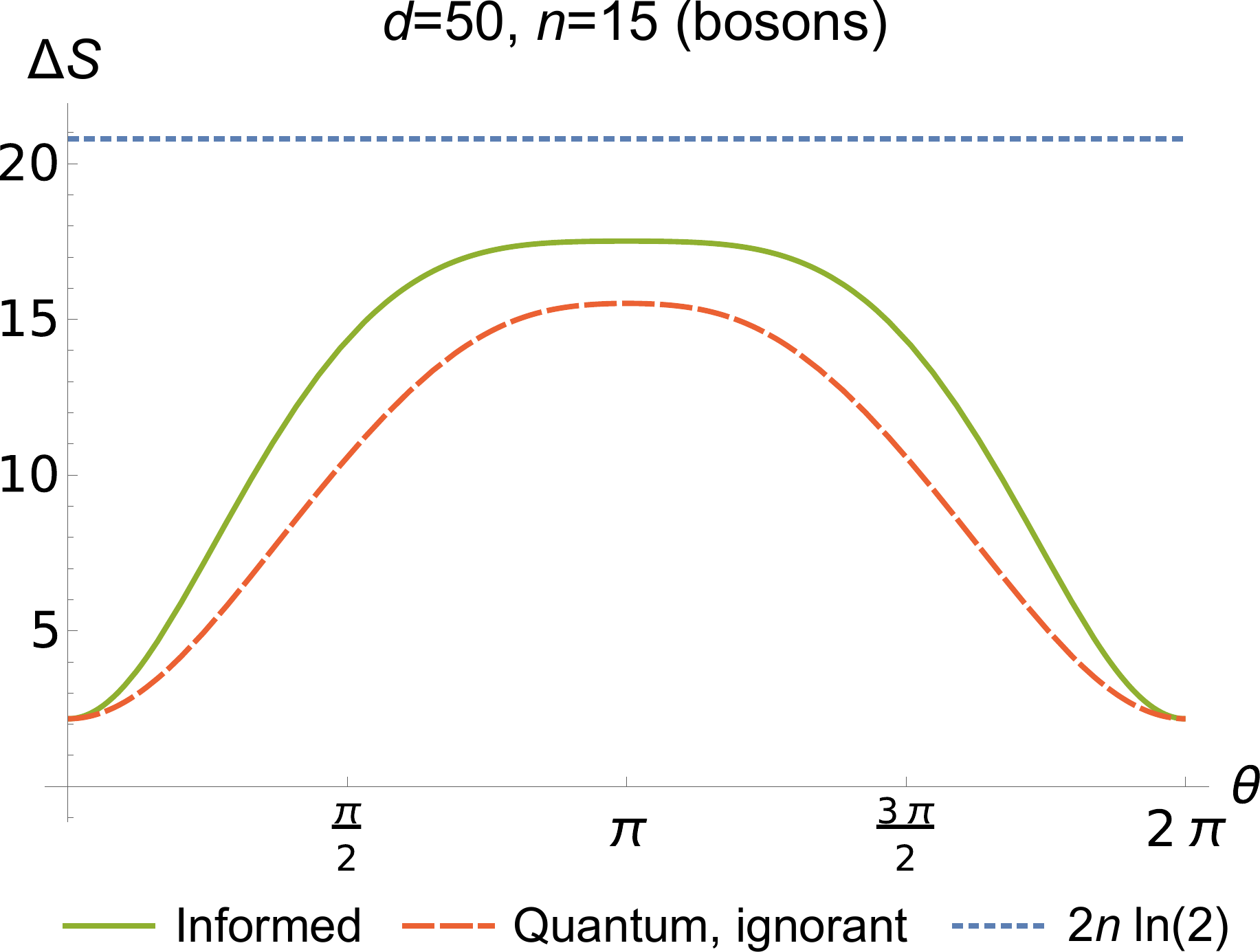} 
                \caption{\textbf{Results for partially distinguishable spins.} {Plots of $\DS{info},\DS{igno}$ as a function of orthogonality of the spin states as determined by $\theta$ in \eqref{eqn:rotated_spin}. The figure is for a bosonic system with initial numbers of particles on either side of the box $n=m=15$, and $d=50$ cells. For comparison, the figure also displays the classical change in entropy, $2n \ln (2)$. Here, the greatest change in entropy occurs when the spin states are orthogonal at $\theta=\pi$.} \label{fig:partial_plot} }
        \label{fig:partial_plot}
\end{figure}

\section*{Discussion}
In contrast to the classical Gibbs paradox setting, we have shown that quantum mechanics permits the extraction of work from apparently indistinguishable gases, without access to the degree of freedom that distinguishes them.
It is notable that the lack of information about this ``spin" does not in principle impede an experimenter at all in a suitable macroscopic limit with large particle number and low density -- the thermodynamical value of the two gases is as great as if they had been fully distinguishable.

The underlying mechanism is a generalisation of the famous Hong-Ou-Mandel (HOM) effect in quantum optics~\cite{Hong1987Measurement,adamson2008detecting,Stanisic2018Discriminating}.
In this effect, polarisation may play the role of the spin.
Then a non-polarising beam splitter plus photon detectors are able to detect whether a pair of incoming photons are similarly polarised.
The whole apparatus is polarisation-independent and thus accessible to the ignorant observer.
Given this context, it is therefore not necessarily surprising that quantum Gibbs mixing can give different results to the classical case.
However, the result of the low density limit is not readily apparent.
This limit is reminiscent of the result in quantum reference frame theory~\cite{bartlett2007reference} that the lack of a shared reference frame presents no obstacle to communication given sufficiently many transmitted copies~\cite{Bartlett2003Classical}.

Two recent papers~\cite{holmes2020enhanced,holmes2020Gibbs} have studied Gibbs-type mixing in the context of optomechanics.
A massive oscillator playing the role of a work reservoir interacts with the photons via their pressure. This oscillator simultaneously acts as a beam splitter between the two sides of the cavity.
In Ref.~\cite{holmes2020enhanced}, the beam splitter is non-polarising and thus (together with the interaction with the oscillator) accessible to the ignorant observer.
The main behaviour there is driven by the HOM effect, which enhances the energy transfer to the oscillator, albeit in the form of fluctuations.
In Ref.~\cite{holmes2020Gibbs}, which studies Gibbs mixing as a function of the relative polarisation rotations between left and right, bosonic statistics are therefore described as acting oppositely to Gibbs mixing effects -- which is different from our conclusions.
However, there is no contradiction: we have shown that an advantage is gained by optimising over all allowed dynamics.
Moreover, the scheme in Ref.~\cite{holmes2020Gibbs} uses a polarisation-dependent beam splitter, which is only accessible to the informed observer.
Therefore the effect described here cannot be seen in such a set-up.
It is an interesting question whether such proposals can be modified to see an advantage of the type described here, even if not optimal.

It is important to determine how the thermodynamic enhancements predicted in this paper may have implications for physical systems. Such an investigation should make use of more practical proposals (such as Refs.~\cite{holmes2020enhanced,holmes2020Gibbs,Myers2020Bosons} ) to better understand possible realisations of mixing. For example, systems of ultra-cold atoms in optical lattices \cite{kaufman2016quantum} may provide a suitable platform to experimentally realise the thermodynamic effects predicted in this work.
The question of the maximal enhancement in the macroscopic limit is particularly compelling given the rapid progress in the manipulation of large quantum systems~\cite{Frowis2018Macroscopic}.

\section*{Methods}
The Supplementary Information contains detailed proofs.
Supplementary Note 1 describes the treatment of classical particles, starting from a description akin to first-quantisation, and then coarse-graining the state space along with appropriate restrictions on the allowed dynamics.
Supplementary Note 2 fills out the derivation for the quantum ignorant observer sketched in the main text.
Supplementary Note 3 provides details for the general case of non-orthogonal spins.
Supplementary Note 4 computes the dimensions of the spaces $\mc{H}_x^\lambda$ from representation theory formulas.
Supplementary Notes 5 and 6 show how to take the low density and large particle number limits, respectively.

\begin{acknowledgments}
We acknowledge financial support from the European Research Council (ERC) under the Starting Grant GQCOP (Grant No.~637352), the EPSRC (Grant No.~EP/N50970X/1) and B.M's doctoral prize EPSRC grant (EP/T517902/1). B.Y.\ is also supported by grant number (FQXi FFF Grant number FQXi-RFP-1812) from the Foundational Questions Institute and Fetzer Franklin Fund, a donor advised fund of Silicon Valley Community Foundation. We are grateful to Zo\"e Holmes, Gabriel Landi, Vlatko Vedral, Chiara Marletto, Bernard Kay, Matthew Pusey, Alessia Castellini and Felix Binder for valuable discussions.
\end{acknowledgments}

\section*{Author contributions}
B.Y. and B.M. contributed equally to the research and writing the manuscript, supervised by G.A.

\section*{Code availability}
Source code for generating the plots is available from the authors upon request.

\section*{Data availability}
No data sets were generated or analysed.

\newpage
\appendix
\onecolumngrid

\section*{Supplementary Note 1 -- Classical treatment}\label{app:classical_treatment}
\subsection*{Classical state space and microscopic dynamics}
Here, we describe the classical setting with identical particles having an internal spin degree of freedom that is not accessed by the experimenter.
The aim is to give a treatment that parallels the quantum one so that the two cases can be compared fairly.
Each particle has two degree of freedom -- a position $x = 1,\dots,d$ and a spin $s = 1,\dots,S$ -- which are the accessible and hidden degrees of freedom, respectively.
(Note that we only require $S=2$ in the main text.)

We start from the point of view of a hypothetical observer for whom \emph{all} the particles are fully distinguishable.
The effective indistinguishability of the particles will be imposed later by a suitable restriction on the allowed operations.
This is rather like the first-quantised description of quantum identical particles.
The underlying state space of $N$ distinguishable particles is
\begin{equation}
	\Sigma_N = \{ (\bm{x}, \bm{s}) \mid \bm{x} \in [d]^N, \bm{s} \in [S]^N \},
\end{equation}
where $[k] = \{1,2,\dots,k\}$. This can be expressed as a Cartesian product $\Sigma_N = \Sigma_N^x \times \Sigma_N^s$ of the individual spaces for each degree of freedom.

A thermodynamical operation involves coupling the particles to a heat bath and work reservoir, the latter two of which we group into a joint system called the ``apparatus'' $A$.
This has its own state space $\Sigma_A$ whose states we designate by a label $a$.
A state of the whole system can therefore be specified by a tuple $(\bm{x}, \bm{s}, a)$.
We assume the underlying microscopic dynamics to be deterministic and reversible; thus, an evolution of whole system consists of an invertible mapping
\begin{equation}
	(\bm{x}, \bm{s}, a) \to (\bm{x}', \bm{s}', a').
\end{equation}

\subsection*{Dynamics independent of spin and particle label}
Now we impose the condition that the operation be spin independent.
This translates into two features: i) the spins are all unchanged, so $\bm{s}' = \bm{s}$, and ii) $\bm{x'}$ and $a'$ are functions of $\bm{x}$ and $a$ only, not $\bm{s}$.
It is clear that $\bm{s}$ is completely decoupled from the other variables, so that the dynamics of the apparatus are the same for any value of $\bm{s}$.
Thus we can drop the redundant information and designate states of the whole system by $(\bm{x}, a)$.

Next, we impose operational indistinguishability of the particles, again by restricting the allowed operations.
An allowed operation must be invariant under a rearrangement of particle labels.
For a permutation $\pi \in S_N$, let $\pi[\bm{x}] = (x_{\pi(1)},\dots, x_{\pi(N)})$.
Then we require that
\begin{equation} \label{eqn:permutation_commuting}
	(\bm{x},a) \to (\bm{x}',a') \Rightarrow (\pi[\bm{x}], a) \to (\pi[\bm{x}'], a') \quad \forall \pi \in S_N,
\end{equation}
i.e., the transformation commutes with all permutations.
This condition implies that $a'$ is a function only of $a$ and the \emph{type} $\bm{t}$ of $\bm{x}$.
By this, we mean $\bm{t} = (t_1,\dots,t_d)$ specifies the number $t_i$ of particles in each cell $i$.
It is then clear that, as far as the dynamics of $A$ are concerned, it is sufficient to keep track of just $(\bm{t}, a)$.
The total number of effective microstates of the particles, as seen by the ignorant observer, is then the number of possible types, equal to $\binom{N+d-1}{N}$.

\subsection*{Subtlety with overly constrained dynamics}
However, there is a subtlety: one can ask whether all (deterministic and reversible) dynamics in the space of $(\bm{t}, a)$ are possible under the constraint Eq.~\eqref{eqn:permutation_commuting}.
If $(\bm{t},a) \to (\bm{t'},a')$ is possible, then there exist some $\bm{x},\bm{x}'$ of types $\bm{t},\bm{t}'$ respectively such that $(\bm{x},a) \to (\bm{x}',a')$.
The condition Eq.~\eqref{eqn:permutation_commuting} then determines how all the remaining vectors $\pi[\bm{x}]$ of type $\bm{t}$ evolve.
There may be a contradiction here -- there are two ways in which a transformation might not be possible:
\begin{itemize}
	\item If there exists $\pi$ such that $\pi[\bm{x}] = \bm{x}$ but $\pi[\bm{x}'] \neq \bm{x}'$, then the transformation cannot be deterministic.
	\item If there exists $\pi$ such that $\pi[\bm{x}] \neq \bm{x}$ but $\pi[\bm{x}'] = \bm{x}'$, then the transformation cannot be reversible.
\end{itemize}
We give the following example, consider $\bm{x}=(1,1),\, \bm{x}'=(1,2)$, which have types $\bm{t}=(2,0),\, \bm{t'}=(1,1)$.
A swap of the two particles preserves $\bm{x}$ but not $\bm{x}'$ -- it is clear that a transition $\bm{t} \to \bm{t'}$ cannot be possible.
In other words, this is because there is no way of ``picking out'' a particle from cell 1 and moving it to cell 2 in a way that acts non-preferentially on the particles.
In quantum mechanics, this obstacle is avoided because it is possible to act symmetrically on the particles such that the final state is an equal superposition of the two $\bm{x}'=(1,2)$ and $(2,1)$.


This hints at a way to avoid the problem in the classical case: widening the scope to include stochastic operations.
Since it is crucial to require that all dynamics are microscopically deterministic, we introduce stochasticity using additional degrees of freedom containing initial randomness.
These couple to the different ways the particles can be permuted, and must necessarily be \emph{hidden}, i.e., not accessible to the observer, in order to maintain ignorance about the particle labels.
The idea is to construct globally deterministic, reversible dynamics such that tracing out the hidden degrees of freedom gives stochastic dynamics on $(\bm{x},a)$ via the probabilities $p(\bm{x}',a'|\bm{x},a)$.
Analogously to Eq.~\eqref{eqn:permutation_commuting}, we impose the condition
\begin{equation} \label{eqn:stoch_perm_commuting}
	p(\bm{x'},a'|\bm{x},a) = p(\pi[\bm{x'}],a'|\pi[x],a) \quad \forall \pi \in S_N.
\end{equation}

The claim is that such dynamics exist that enable all possible (deterministic, reversible) transformations of $(\bm{t},a)$.
To see this, consider just one desired transformation $(\bm{t},a) \to (\bm{t'},a')$.
We introduce two sets of additional variables $\bm{h_1},\bm{h_2}$ which respectively contain information about $\bm{x}$ and $\bm{x'}$.
$\bm{h_1}$ starts in a ``ready'' state $\bm{0}$, while $\bm{h_2}$ is uniformly distributed over all $\bm{x}'$ of type $\bm{t'}$.
Writing a joint state of all subsystems as $(\bm{x},a,\bm{h_1},\bm{h_2})$, it is easily verified that the following dynamics are deterministic and reversible:
\begin{equation}
	(\bm{x},a,\bm{0},\bm{x'}) \to (\bm{x'},a',\bm{x},\bm{x'}) \quad \forall \bm{x},\bm{x'} \text{ of types } \bm{t},\bm{t}',
\end{equation}
where $a'$ is of course a function of $\bm{t}$ only.
Here, $\bm{h_1}$ keeps a record of the initial configuration (to ensure reversibility) and $\bm{h_2}$ randomises the final configuration to range uniformly over all $\bm{x'}$ of type $\bm{t'}$.
Hence we see that $p(\bm{x}',a'|\bm{x},a)$ is constant over all $\bm{x},\bm{x'}$ of interest and thus satisfies condition Eq.~\eqref{eqn:stoch_perm_commuting}.

Note that $\bm{h_1}$ has to be initialised in a ``pure'' state of zero entropy such that it can record information.
Such a state, being non-thermal, should be regarded as an additional resource which costs work to prepare.
(By contrast, the uniformly random variable $\bm{h_2}$ is thermal and thus free.)
The necessary leakage of information into $\bm{h_1}$ therefore entails dissipation of work into heat.
Hence the work extraction formula~\eqref{eqn:class_indist} is technically an upper bound to what can be achieved classically.

This record of information about the initial configuration is seen to be necessary only for those transitions where the set of $\bm{x}$ of type $\bm{t}$ is smaller than the set of $\bm{x'}$ of type $\bm{t'}$, in order to prevent irreversible merging of states.
This situation can be avoided, for instance, in the case of the classical analogue of fermions wherein no more than one particle can occupy a cell.
Similarly, in the low density limit (discussion of which appears in the main text), almost all configurations are of this type with very high probability. (One could also argue that this problem is never encountered in reality -- as soon as two particles overlap sufficiently, we are already in the quantum parameter regime.)

To summarise what we have shown in this section:
\begin{itemize}
	\item Classical identical particles can be treated, analogously to the quantum case, as (in principle) distinguishable particles whose dynamics are restricted to be independent of particle label.
	\item An observer with access only to spin-independent operations can treat the system as if the particles were spin-less.
	\item There is a subtlety with the particle-label-independent operations that blocks certain transitions. This restriction can be lifted with additional degrees of freedom but may require dissipation of work into heat. This extra cost is zero when particles always occupy distinct cells.
\end{itemize}


\section*{Supplementary Note 2 -- Details for quantum ignorant observer} \label{app:quantum_details}
In this section, we provide additional details for the entropy change as seen by the ignorant observer.\\

Recall that Schur-Weyl duality~\cite[Chapter 5]{harrow2005applications} provides the decomposition
\begin{equation}
	\mc{H}_x^{\ox N} = \bigoplus_\lambda \mc{H}_x^\lambda \ox \mc{K}_x^\lambda,
\end{equation}
where $\lambda$ runs over all Young diagrams containing $N$ boxes and no more than $d$ rows.
A Young diagram $\lambda$ is a set of unlabelled boxes arranged in rows, with non-increasing row length from top to bottom.
We can equivalently describe $\lambda=(\lambda_1,\lambda_2,\dots,\lambda_d)$, where $\lambda_i$ is the number of boxes in row $i$.
\ytableausetup{smalltableaux}
For example,~ \ydiagram{3,1}~would be denoted $(3,1)$ (where $N=4,d=2$).

$\mc{H}_x^\lambda$ and $\mc{K}_x^\lambda$ carry irreps of $\mathrm{U}(d)$ and $S_N$ respectively, corresponding to irreducible subspaces under the actions of single-particle unitary rotations $u^{\ox N} \ox I^{\ox N}$ and particle label permutations $\Pi \ox I^{\ox N}$, each of which act only on the spatial part.
The same decomposition works for the spin part $\mc{H}_s^{\ox N}$, although now the Young diagrams $\lambda$ have maximally two rows.
In fact, they correspond to the familiar $\mathrm{SU}(2)$ irreps with total angular momentum $J$, via $\lambda = (N/2+J,\, N/2-J)$.

After putting the spatial and spin decompositions together, projecting onto the overall (anti-)symmetric subspace causes the symmetries of the two components to be linked.
For bosons, the overall symmetric subspace (itself a trivial irrep of $S_N$) occurs exactly once in $\mc{K}_x^\lambda \ox \mc{K}_s^{\lambda'}$ if and only if $\lambda = \lambda'$, and otherwise does not~\cite[Section 7-13]{Hamermesh1989}.
Thus we have
\begin{align} \label{eqn:decomp_bosons}
	\mc{H}_N & = \bigoplus_{\lambda,\lambda'} \mc{H}_x^\lambda \ox \mc{H}_s^{\lambda'} \ox P_+ \left[\mc{K}_x ^\lambda \ox \mc{K}_s^{\lambda'} \right] \nonumber \\
		& = \bigoplus_\lambda \mc{H}_x^\lambda \ox \mc{H}_s^\lambda \quad \text{(bosons)}.
\end{align}
For fermions, the only difference is that the projector $P_-$ onto the antisymmetric subspace enforces $\lambda' = \lambda^T$, denoting the transpose of the Young diagram in which rows and columns are interchanged; thus,
\begin{equation} \label{eqn:decomp_fermions}
	\mc{H}_N = \bigoplus_\lambda \mc{H}_x^{\lambda^T} \ox \mc{H}_s^{\lambda} \quad \text{(fermions)}.
\end{equation}
Due to the use of a two-dimensional spin, we employ the correspondence $J \leftrightarrow \lambda=(N/2+J,\, N/2-J, 0,0,\dots)$ (with a total of $d$ rows) to replace the label $\lambda$ by $J$.

Let us first consider the bosonic case.
Thanks to the decomposition in Eq.~\eqref{eqn:decomp_bosons}, a state $\rho$ (as seen by the informed observer) can be written in terms of the basis ${\ket{J,q}}_x {\ket{J,M}}_s {\ket{\phi_J}}_{xs}$, where $\ket{J,q} \in \mc{H}_x^J, \, \ket{J,M} \in \mc{H}_s^J,\, \ket{\phi_J} \in \mc{K}_x^J \ox \mc{K}_s^J$, as described in the main text.
The ignorant observer sees the reduced state after tracing out the spin part, of the form
\begin{align}
	\rho_x & = \tr_s \rho = \bigoplus_J p_J \rho_x^{J} \ox \tr_s {\proj{\phi_J}}_{xs}.
\end{align}
The entropy of this state is
\begin{equation}
	S(\rho_x) = H(\bm{p}) + \sum_J p_J \left[ S\left(\rho_x^{J}\right) + S\left( \tr_s {\proj{\phi_J}}_{xs}\right) \right],
\end{equation}
where $H(\bm{p}) := -\sum_J p_J \ln p_J$ is the Shannon entropy of the probability distribution $p_J$.

As argued in the main text, the fully thermalised final state is of the form
\begin{equation}
	\rho'_x = \bigoplus_J p_J \frac{I^J_x}{d_J} \ox \tr_s {\proj{\phi_J}}_{xs},
\end{equation}
with entropy
\begin{equation}
	S(\rho'_x) = H(\bm{p}) + \sum_J p_J \left[ \ln d_J  + S\left( \tr_s {\proj{\phi_J}}_{xs}\right) \right].
\end{equation}

An example of a channel that achieves the mapping from $\rho_x$ to $\rho'_x$ -- albeit without a coupling to a heat bath or work reservoir -- is the so-called ``twirling'' operation.
This is a probabilistic average over all single-particle unitary rotations $u_x^{\ox N}$:
\begin{align}
	\mc{T}_x(\rho) = \int \dd \mu(u_x) \, u_x^{\ox N} \rho {u_x^{\ox N}}^\dagger ,
\end{align}
where $\mu$ is the Haar measure over the group $\mathrm{U(d)}$. 

The entropy change for the ignorant observer is therefore
\begin{equation} \label{eqn:entropy_change_middle}
	\DS{igno} = S(\rho'_x) - S(\rho_x) = \sum_J p_J \left[ \ln d_J - S\left( \rho_x^{J} \right) \right].
\end{equation}
(Note that the states $\phi_J$ do not enter into the entropy change.)
Our goal is therefore to determine the probabilities $p_J$, dimensions $d_J$, and the entropy of the component states $\rho_x^{J}$.

The case of indistinguishable gases is dealt with in the main text: the state is fully in the subspace $J=N/2$, corresponding to the spatially symmetric subspace for bosons and spatially antisymmetric for fermions.

For gases of different spins, the initial state is such that all particles on the left are in $\ket{\uparrow}$ and all on the right are in $\ket{\downarrow}$.
Before getting to the thermal state, first consider a pure state in which $n_i$ particles are in each cell $i$ on the left, and $m_i$ in each cell $i$ on the right (such that $\sum_i n_i = n,\, \sum_i m_i = m$).
This spatial configuration is denoted by the pair of vectors $(\bm{n},\bm{m})$.
The properly symmetrised wavefunction is
\begin{align} \label{eqn:initial_pure}
	\ket{\psi(\bm{n},\bm{m})} & = \mc{N}(\bm{n},\bm{m}) \sum_{\text{distinct } \pi \in S_N} \pi {\ket{\bm{n},\bm{m}}}_x \ox \pi {\ket{\uparrow^n \downarrow^m}}_s, \nonumber \\
	\ket{\bm{n},\bm{m}} & := \ket{1_L^{n_1} 2_L^{n_2} \dots 1_R^{m_1} 2_R^{m_2} \dots},
\end{align}
where $\pi$ runs over permutations of the $N$ particles that lead to \emph{distinct} terms $\pi {\ket{\bm{n},\bm{m}}}_x$.
(This is well-defined, since whenever $\pi$ and $\pi'$ have the same effect on $\ket{\bm{n},\bm{m}}$, they must also have the same effect on $\ket{\uparrow^n \downarrow^m}$.)
$\mc{N}(\bm{n},\bm{m})$ is a normalisation factor (such that $\mc{N}(\bm{n},\bm{m})^{-2}$ is the number of distinct terms in the sum).
We determine the $p_J$ via the expectation value of the projector $P_s^J$ onto the subspace $\mc{H}_s^J$:
\begin{align} \label{eqn:initial_pure_trace_x}
	& \braXket{\psi(\bm{n},\bm{m})}{P_s^J}{\psi(\bm{n},\bm{m})}  \nonumber \\
	& = \mc{N}(\bm{n},\bm{m})^2 \sum_{\text{distinct } \pi,\pi'} \braXket{\bm{n},\bm{m}}{\pi' \pi}{\bm{n},\bm{m}} \braXket{\uparrow^n \downarrow^m}{\pi' P_s^J \pi}{\uparrow^n \downarrow^m} \nonumber \\
	& = \mc{N}(\bm{n},\bm{m})^2 \sum_{\text{distinct } \pi} \braXket{\uparrow^n \downarrow^m}{\pi P_s^J \pi}{\uparrow^n \downarrow^m},
\end{align}
where the second line holds because any pair of $\pi,\pi'$ giving rise to distinct terms in Eq.~\eqref{eqn:initial_pure} also have different actions on $\ket{\bm{n},\bm{m}}$.
Now we use Clebsch-Gordan coefficients to evaluate each term in this last sum.
First note that we can express $\ket{\uparrow^n}$ as a combined spin with $J_1=M_1=n/2$, and similarly $\ket{\downarrow^m}$ as a spin with $J_2=-M_2=m/2$. The Clebsch-Gordan coefficient $C(\frac{n}{2},\frac{n}{2};\frac{m}{2},\frac{-m}{2}; J, \frac{n-m}{2})$ is precisely the amplitude for this state in the $J$ subspace.
This is unchanged by the inclusion of a permutation $\pi$, so Eq.~\eqref{eqn:initial_pure_trace_x} simplifies to
\begin{align}
	\braXket{\psi(\bm{n},\bm{m})}{P_s^J}{\psi(\bm{n},\bm{m})} = \abs{ C\left(\frac{n}{2},\frac{n}{2};\frac{m}{2},\frac{-m}{2}; J, \frac{n-m}{2}\right) }^2.
\end{align}
Now it remains to consider the correct initial state, which is a uniform probabilistic mixture of all $\ket{\psi(\bm{n},\bm{m})}$ with a fixed number of particles $n,m$ on the left and right, respectively.
Since the Clebsch-Gordan coefficient is the same for all such configurations, we have~\cite{BiedenharnLouck}
\begin{align} \label{eqn:pj_cg_coeff}
	p_J & = \abs{ C\left(\frac{n}{2},\frac{n}{2};\frac{m}{2},\frac{-m}{2}; J, \frac{n-m}{2}\right) }^2 \nonumber \\
		& = \frac{(2J+1) n! m!}{\left(\frac{N}{2}+J+1\right)! \left(\frac{N}{2}-J\right)!}.
\end{align}

Finally, we determine the entropy of each $\rho_x^J$ component.
Using the basis ${\ket{J,q}}_x {\ket{J,M}}_s {\ket{\phi_J}}_{xs}$ provided by the Schur-Weyl decomposition, we have
\begin{align} \label{eqn:pure_schur_decomp}
	\ket{\psi(\bm{n},\bm{m})} = \sum_J \sqrt{p_J} {\ket{\psi(\bm{n},\bm{m},J)}}_x {\ket{J,\frac{n-m}{2}}}_s {\ket{\phi_J}}_{xs}.
\end{align}
Here, ${\ket{\psi(\bm{n},\bm{m},J)}}_x \in \mc{H}_x^J$ is some linear combination of the ${\ket{J,q}}_x$ -- without needing to determine these states entirely, it will be sufficient to note that they are orthogonal for different configurations $(\bm{n},\bm{m})$. This follows from the fact that different $\ket{\psi(\bm{n},\bm{m})}$ are fully distinguishable just by measuring the occupation numbers of different cells. Tracing out $s$, we find
\begin{align}
	\tr_s \psi(\bm{n},\bm{m}) & = \bigoplus_J p_J \psi(\bm{n},\bm{m},J) \ox \tr_s {\proj{\phi_J}}_{xs}, \nonumber \\
	\rho_x^J & \propto  \sum_{\bm{n},\bm{m}} \psi(\bm{n},\bm{m},J).
\end{align}
From orthogonality of the $\psi(\bm{n},\bm{m},J)$, it follows that
\begin{align} \label{eqn:initial_entropy}
	S(\rho_x^J) = \ln \binom{n+d/2-1}{n} + \ln \binom{m+d/2-1}{m}.
\end{align}
Inserted into Eq.~\eqref{eqn:entropy_change_middle}, this results in the claimed entropy changes~\eqref{eqn:entropy_bosons}, \eqref{eqn:entropy_fermions}.

{
\section*{Supplementary Note 3 -- Partial distinguishability}
Here, we extend the analysis to include non-orthogonal spins states.
As before, we keep the initial spins on the left side of the box as ${\ket{\uparrow}}^{\ox n}$, but now on the right we have ${\ket{\nearrow}}^{\ox m}$, where $ \ket{\nearrow} = \cos(\theta/2) \ket{\uparrow} + \sin(\theta/2) \ket{\downarrow}$.

\subsection*{Informed observer}
Let us first discuss the operations allowed to be performed by the informed observer.
They are permitted to know about the value of the spins in the $\ket{\uparrow}, \ket{\downarrow}$ basis; they may engineer dynamics diagonal in this basis.
Of course, this choice entails a preferred spin basis -- this is necessary in order to have a well-defined notion of conditioning dynamics on the value of a spin.
We thus require a global unitary of the form $U = \bigoplus_M U^{(M)}_{xsBW}$, where the block structure refers to subspaces with fixed $M$ as defined by the Schur basis.
Under a block-diagonal operation, one cannot extract work from coherences between the blocks~\cite{horodecki2013fundamental,Lostaglio2015Description}; that is, the initial state of the spins can be effectively replaced by the dephased state
\begin{equation}
	\Phi(\rho_{xs}) := \sum_M Q^M_s \rho_{xs} Q^M_s = \sum_M q_M \rho^{(M)}_{xs},
\end{equation}
where $Q_s^M$ is the projector onto the $M$ block.
In other words, the state behaves thermodynamically as a statistical mixture of the different $z$-spin numbers $M$.
It follows that the overall entropy change is the average

	\begin{equation}
		\DS{info}(\rho_{xs}) = \sum_M q_M \DS{info}(\rho_{xs}^{(M)}).
	\end{equation}

As for the case of orthogonal spins, the initial state is a uniform mixture of states generalising equation Eq.~\eqref{eqn:initial_pure},
\begin{equation} \label{eqn:pure_nonorthogonal}
	\ket{\psi(\bm{n},\bm{m})} = \mc{N}(\bm{n},\bm{m}) \sum_{\text{distinct } \pi \in S_N} \pi {\ket{\bm{n},\bm{m}}}_x \ox \pi {\ket{\uparrow^n \nearrow^m}}_s,
\end{equation}
where again it is sufficient (and well-defined) for $\pi$ to run only over permutations that lead to distinct $\pi {\ket{\bm{n},\bm{m}}}_x$.
As before, $\mc{N}^{-2}$ is simply the number of such distinct terms (independent of $\theta$).

Expanding $\ket{\nearrow^m}$ in the preferred basis, it is easily seen that

	\begin{equation}
		\ket{\nearrow^m} = \sum_{k=0}^m \cos(\theta/2)^{m-k} \sin(\theta/2)^k \sum_{\text{distinct } \pi \in S_m} \pi \ket{\uparrow^{m-k} \downarrow^k},
	\end{equation}

and so

	\begin{equation}\label{eqn:qM_expansion}
		q_M = \braXket{\uparrow^n \nearrow^m}{Q^M_s}{\uparrow^n \nearrow^m} = \binom{m}{(n+m)/2-M} \cos(\theta/2)^{m-n+2M} \sin(\theta/2)^{n+m-2M},
	\end{equation}

having used $M = (n+m)/2-k$.
Without needing to know the form of $Q^M_s \ket{\psi(\bm{n},\bm{m})}$, it is sufficient to note that all such states are pure and must be orthogonal, since they can be mutually perfectly distinguished by measuring the occupation number in each cell.
The entropy $S(\rho^{(M)}_{xs})$ is therefore just as in Eq.~\eqref{eqn:initial_entropy} for each $M$.

Due to the block-diagonal structure of the global unitary $U$, the maximum entropy final state is given by a maximally mixed state for each $M$ block.
Considering the number of possible spatial configurations for a fixed number of up and down spins, the dimension of the $M$ block is found to be $\binom{(n+m)/2-M+d-1}{(n+m)/2-M} \binom{(n+m)/2+m+d-1}{(n+m)/2+M}$ in the bosonic case.
Hence the overall entropy change is

	\begin{equation}
		\DS{info} = \sum_{M=-N/2}^{N/2} q_M \left[ \ln \binom{N/2-M+d-1}{N/2-M} + \ln \binom{N/2+M+d-1}{N/2+M} \right] - \left[ \ln \binom{n+d/2-1}{n} + \ln \binom{m+d/2-1}{m} \right].
	\end{equation}

In the fermionic case, analogous counting gives

	\begin{equation}
		\DS{info} = \sum_{M=-N/2}^{N/2} q_M \left[ \ln \binom{d}{N/2-M} + \ln \binom{d}{N/2+M} \right] - \left[ \ln \binom{d/2}{n} + \ln \binom{d/2}{m} \right].
	\end{equation}

\subsection*{Ignorant observer}
For the ignorant observer, we now have to analyse $\rho_x^J$.
From Eq.~\eqref{eqn:pure_nonorthogonal} (recalling that the permutations to be summed over are those that lead to distinct $\pi{\ket{\bm{n},\bm{m}}}_x$),

	\begin{align} \label{eqn:nonorthogonal_traced}
		\tr_s \left[ P^J_s \proj{\psi(\bm{n},\bm{m})} \right] & = \mc{N}^2 \sum_{\pi,\pi'} \braXket{\uparrow^n \nearrow^m}{\pi^\dagger P^J_s\pi'}{\uparrow^n \nearrow^m} \pi' \proj{\bm{n},\bm{m}} \pi^\dagger \nonumber \\
			& = \mc{N}^2 \sum_{\pi,\pi'} \braXket{\uparrow^n \nearrow^m}{P^J_s \pi^\dagger \pi'}{\uparrow^n \nearrow^m} \pi' \proj{\bm{n},\bm{m}} \pi^\dagger,
	\end{align}

using the fact that the projector $P^J_s$ commutes with permutations.
In order to simplify this, we examine coefficients of the form $\braXket{\uparrow^n \nearrow^m}{P^J_s \pi}{\uparrow^n \nearrow^m}$.
Using the Schur basis just for the spin part, in general one can expand $\ket{\uparrow^n \nearrow^m} = \sum_{J,M,r} \omega_{J,M,r} \ket{J,M,r}$.
The $r$ label, representing the part of the basis acted upon by the permutation group, consists of any quantum numbers needed to complete the set along with $J$ and $M$.
We describe a convenient choice of such numbers, denoted $j_1, j_{1,2}, \dots, j_{1,\dots,n}$ and $k_1, k_{1,2}, \dots, k_{1,\dots,m}$.
$j_1$ is the total spin eigenvalue of spin 1, $j_{1,2}$ of spins 1 and 2 together, and so on.
$k_1,\dots$ have the same meaning, but for the remaining spins label $n+1,\dots,n+m$.
That these complete the set of quantum numbers is evident from imagining performing an iterated Clebsch-Gordan procedure.
This would involve coupling spins 1 and 2, then adding in spin 3, and so on up to spins $n$.
Spins $n+1$ up to $n+m$ would be coupled recursively in the same manner, and then finally the two blocks of spins coupled to give the overall $J$.

For the state $\ket{\uparrow^n \nearrow^m}$ each of the two blocks of spins is fully symmetric, meaning that each of these spin eigenvalues is maximal: $j_1 = k_1 = \frac{1}{2},\, j_2 = k_2 = 1,\, \dots, \ j_{1,\dots,n} = \frac{n}{2},\, k_{1,\dots,m} = \frac{m}{2}$.
Given this choice of basis, there is only a single value of $r=r_0$ in the expansion of $\ket{\uparrow^n \nearrow^m}$, referring to this collection of spin eigenvalues.
Therefore we can write $\ket{\uparrow^n \nearrow^m} = \sum_{J,M} \omega_{J,M} \ket{J,M,r_0}$, and

	\begin{align}
		\braXket{\uparrow^n \nearrow^m}{P^J_s \pi}{\uparrow^n \nearrow^m} & = \sum_{M,M'} \omega_{J,M'}^* \omega_{J,M} \braXket{J,M',r_0}{\pi}{J,M,r_0} \nonumber \\
			& = \sum_M \abs{\omega_{J,M}}^2 \braXket{J,M,r_0}{\pi}{J,M,r_0} \nonumber \\
			& =: \sum_M \abs{\omega_{J,M}}^2 \eta_J(\pi)
	\end{align}

since $\braXket{J,M,r_0}{\pi}{J,M,r_0}$ is independent of $M$ (and $r_0$ is fixed anyhow).
Expanding $\ket{\uparrow^n \nearrow^m}$ in the preferred basis and using the Clebsch-Gordan coefficients for coupling the two blocks of spins gives

	\begin{equation}
		\abs{\omega_{J,M}}^2 = q_M \abs{C\left( \frac{n}{2},\frac{n}{2}; \frac{m}{2},M-\frac{n}{2}; J, M \right)}^2,
	\end{equation}

where $q_M$ is defined in Eq.~\eqref{eqn:qM_expansion}.
Putting this into Eq.~\eqref{eqn:nonorthogonal_traced}, we have

	\begin{align}
		\tr_s \left[ P^J_s \proj{\psi(\bm{n},\bm{m})} \right] & = \mc{N}^2 \sum_{\pi,\pi'} \sum_M \abs{\omega_{J,M}}^2 \eta_J(\pi^\dagger \pi') \pi'\proj{\bm{n},\bm{m}} \pi^\dagger \nonumber \\
			& = \sum_M q_M \abs{C\left( \frac{n}{2},\frac{n}{2}; \frac{m}{2},M-\frac{n}{2}; J, M \right)}^2  \left[ \mc{N}^2 \sum_{\pi,\pi'} \eta_J(\pi^\dagger \pi') \pi' \proj{\bm{n},\bm{m}} \pi^\dagger \right].
	\end{align}

Crucially, the state in brackets is independent of $M$ and must therefore be identical to the state we named $\ket{\psi(\bm{n},\bm{m},J)}$ in Eq.~\eqref{eqn:pure_schur_decomp}.
Hence the remaining analysis runs exactly as in the orthogonal spin case, apart from the replacement of $p_J$ by $\sum_M q_M \abs{C\left( \frac{n}{2},\frac{n}{2}; \frac{m}{2},M-\frac{n}{2}; J, M \right)}^2$.
Thus, all that changes is the probability distribution over $J$, and this only depends on the probability over $M$, determined ultimately by the angle $\theta$.
}

\section*{Supplementary Note 4 -- Dimension counting} \label{app:dimensions}
From~\cite[Chapter 7]{GoodmanWallach}, we have (labelling by $\lambda$ instead of $J$)
\begin{align} \label{eqn:unitary_rep_dimension}
	\dim \mc{H}_x^\lambda & = \frac{\prod_{1\leq i < j \leq d} (\tilde{\lambda}_i-\tilde{\lambda}_j)}{\prod_{m=1}^{d-1} m!}, \nonumber \\
	\tilde{\lambda} & := \lambda + (d-1, d-2, \dots, 0).
\end{align}
First take the bosonic case.
Since the Young diagram for the $\mathrm{SU}(2)$ spin representation has no more than two rows, the same $\lambda$ labelling the spatial part has no more than two \emph{non-zero} rows.
Hence we have $\tilde{\lambda} = \left( \frac{N}{2}+J+d-1,\, \frac{N}{2}-J+d-2,\, d-3,\, d-4, \dots 0 \right)$. Calculating the product in the numerator of Eq.~\eqref{eqn:unitary_rep_dimension} is aided by the table below, which lists the values of $\tilde{\lambda}_i - \tilde{\lambda}_j$, where $i$ labels the row and $j>i$ labels the column:
\begin{equation}
	\begin{array}{c !{\vline width 1pt} c|c|c|c|c|c|c}
        & 2		& 3		& 4		& 5		& \dots	& d-1	& d		 	\\ \btrule{1pt}
        1		& 2J+1	& \frac{N}{2}+J+2	& \frac{N}{2}+J+3	& \dots	& \dots	& \dots	& \frac{N}{2}+J+d-1 	\\ \hline
        2		& 		& \frac{N}{2}-J+1	& \frac{N}{2}-J+2	& \dots	& \dots & \dots & \frac{N}{2}-J+d-2 	\\ \hline
        3		& 		&		& 1		& 2		& \dots & \dots	& d-3		\\ \hline
        4		& 		& 		&		& 1		& \dots	& \dots	& d-4		\\ \hline
        \vdots	&		&		&		&		&		&		& \vdots	\\ \hline
        d-2		&		&		&		& 		& 		& 1		& 2			\\ \hline
        d-1		&		&		&		&		&		&		& 1
\end{array}
\end{equation}
The product of the terms in the first row is
\begin{equation}
	(2J+1) \frac{(\frac{N}{2}+J+d-1)!}{(\frac{N}{2}+J+1)!},
\end{equation}
the second row gives
\begin{equation}
	\frac{(\frac{N}{2}-J+d-2)!}{(\frac{N}{2}-J)!},
\end{equation}
and the remaining rows give
\begin{equation}
	\prod_{m=1}^{d-3} m!.
\end{equation}
Putting these into Eq.~\eqref{eqn:unitary_rep_dimension} results in the expression for $d^B_{N,J}$ in~\eqref{eqn:dimensions}.

For fermions, we instead use the transpose of the Young diagram, with
\begin{equation}
	\lambda^T = (\underbrace{2,\dots,2}_{\frac{N}{2}-J},\, \underbrace{1,\dots,1}_{2J} ).
\end{equation}
An important restriction on $\lambda^T$ is that the number of rows can never be greater than the dimension, so $\frac{N}{2}+J \leq d$. We find
\begin{equation}
	\tilde{\lambda^T} = ( \underbrace{d+1,d,d-1, \dots, d-\frac{N}{2}+J+2}_{\frac{N}{2}-J},\, \underbrace{d-\frac{N}{2}+J, d-\frac{N}{2}+J-1, \dots, d-\frac{N}{2}-J+1}_{2J}, \underbrace{d-\frac{N}{2}-J-1,\dots,0}_{d-\frac{N}{2}-J}).
\end{equation}
As before, the differences $\tilde{\lambda^T}_i - \tilde{\lambda^T}_j$ can be arranged as follows:\\
\begin{equation}
\scalebox{0.85}{$
\begin{array}{c !{\vline width 1pt} c|c|c|c!{\color{blue}\vline width .8pt} c|c|c|c !{\color{red}\vline width .8pt}  c|c|c|c|c}
            & 2		& 3			    & \dots	    & \frac{N}{2}-J	& \frac{N}{2}-J+1 & \frac{N}{2}-J+2	& \dots   & \frac{N}{2}+J        &\frac{N}{2}+J+1       &\frac{N}{2}+J+2  &\dots & d-1  & d	 	\\ \btrule{1pt}
        1		& 1	    & 2	    	    & \dots		& \frac{N}{2}-J-1	& \frac{N}{2}-J+1 & \frac{N}{2}-J+2 &\dots    & \frac{N}{2}+J        &\frac{N}{2}+J+2       &\frac{N}{2}+J+3	 &\dots &d   &d+1  \\ \hline
        2	    & 	    & 1	    	    & \dots		& \frac{N}{2}-J-2	& \frac{N}{2}-J   & \frac{N}{2}-J+1 &\dots    & \frac{N}{2}+J-1      &\frac{N}{2}+J+1 	  &\frac{N}{2}+J+2  &\dots &d-1   &d\\ \hline
        \vdots	& 	    & 	     	    & 	        & \vdots&\vdots & \vdots&         &\vdots     &\vdots     &\vdots&      &\vdots&\vdots  	\\ \hline
        \frac{N}{2}-J-1	& 	    & 	     	    &   	 	& 1	    & 3     & 4     &\dots    &2J+2      &2J+4      &2J+5 &\dots &d-(\frac{N}{2}-J)+2&d-(\frac{N}{2}-J)+3  	\\ \hline 
        \frac{N}{2}-J		& 	    & 	     	    &   	 	& 	    & 2     & 3     &\dots    &2J+1      &2J+3      &2J+4 &\dots &d-(\frac{N}{2}-J)+1&d-(\frac{N}{2}-J)+2  	\\ \arrayrulecolor{blue}\btrule{.8pt}
        \frac{N}{2}-J+1	& 	    & 	     	    &   	 	& 	    &       &  1    &\dots    &2J-1      &2J+1      &2J+2 &\dots &d-(\frac{N}{2}-J)-1&d-(\frac{N}{2}-J) \\ \arrayrulecolor{black}\hline
        \vdots  & 	    & 	     	    &   	   	&   	&       &       &         & \vdots    & \vdots    &\vdots&      &\vdots&\vdots	\\ \hline
        \frac{N}{2}+J-1    & 	    & 	     	    &   	   	&   	&       &       &         & 1         & 3         &4     &\dots &d-(\frac{N}{2}+J)+1&d-(\frac{N}{2}+J)+2	\\ \hline
        \frac{N}{2}+J      & 	    & 	     	    &   	   	&   	&       &       &         &           & 2         &3     &\dots &d-(\frac{N}{2}+J)&d-(\frac{N}{2}+J)+1	\\ \arrayrulecolor{red}\btrule{.8pt}
        \frac{N}{2}+J+1    & 	    & 	     	    &   	   	&   	&       &       &         &           &           &1     &\dots &d-(\frac{N}{2}+J)-2&d-(\frac{N}{2}+J)-1	\\\arrayrulecolor{black} \hline
        \vdots	& 	    & 	     	    &   	 	&    	&       &       &         &           &           &      &      &\vdots&\vdots	\\ \hline
        d-2	    & 	    & 	     	    &   	 	&    	&       &       &         &           &           &      &      &1     &2	\\ \hline
        d-1	    & 	    & 	     	    &   	    &   	&       &       &         &           &  	      &      &      &      &1
    \end{array}$}
\end{equation}\\
Here, the blue and red lines indicate the division into the three main index groups.
We want to calculate the product of all rows in the table.
The bottom group of rows gives
\begin{equation}
	\prod_{m=1}^{d-(N/2+J)-1} m! .
\end{equation}
The next group up, being careful to discount the terms lost due to the jump at column $j = N/2+J$, gives
\begin{equation}
	\prod_{m=d-(N/2+J)+1}^{d-(N/2-J)} \frac{m!}{m-(d-(N/2+J))}
\end{equation}
Finally, the top group of rows, noting the additional jump at $j = N/2-J+1$, gives
\begin{equation}
	\prod_{m=d-(N/2-J)+2}^{d+1} \frac{m!}{[m-(d-(N/2+J))] [m-(d-(N/2-J)+1)]}.
\end{equation}
Inserting into Eq.~\eqref{eqn:unitary_rep_dimension}, we need to divide the product of the above three terms by $\prod_{m=1}^{d-1}m!$.
This factor cancels all the factorials present in the above three expressions, with the exception of the top two rows, and contributes two factorials occurring at $m = d(N/2+J),\, d-(N/2-J)+1$.
Therefore we have
\begin{align}
	d^F_{N,J} & =  \prod_{r=d-N/2-J+1}^{d-N/2+J} \frac{1}{r-d+N/2+J}  \nonumber \\
		& \quad \cdot \prod_{m=d-N/2+J+2}^{d+1} \frac{1}{(m-d+N/2+J)(m-d+N/2-J-1)} \cdot \frac{d!(d+1)!}{(d-N/2+J+1)!(d-N/2-J)!} \nonumber \\
		& = \frac{1}{(2J)!} \cdot \frac{(2J+1)!}{(N/2+J+1)!(N/2-J)!} \cdot \frac{d!(d+1)!}{(d-N/2+J+1)!(d-N/2-J)!} \nonumber \\
		& = \frac{(2J+1) d!(d+1)!}{(N/2+J+1)!(N/2-J)!(d-N/2+J+1)!(d-N/2-J)!}.
\end{align}

\section*{Supplementary Note 5 -- Low density limit} \label{app:low_density}
\subsection*{Bosons}
Here we prove equation~\eqref{eqn:boson_low_density} for bosons.
For simplicity, we take $n=m$.
The result rests on the observation that, for sufficiently large $d$, the ratio $d^B_J/p_J \approx \binom{n+d-1}{n}^2$.
We have
\begin{align}
    \frac{d^B_J/p_J}{\binom{n+d-1}{n}^2} & = \frac{(d-1)!(d+n+J-1)! (d+n-J-2)!}{(d-2)!(d+n-1)!^2} \nonumber \\
        & = (d-1) \frac{\prod_{k=0}^{J-1}(d+n+k)}{\prod_{k=0}^J(d+n-J-1+k)} \nonumber \\
        & = \left(1-\frac{1}{d}\right) \prod_{k=0}^{J-1}(1+[n+k]/d) \prod_{k=0}^J(1+[n-J-1+k]/d)^{-1}
\end{align}
Letting $x_k = [n+k]/d$, we have
\begin{align}  
    \prod_{k=0}^{J-1}(1+[n+k]/d) & = \sum_{k=0}^{J-1} x_k + \sum_{0=k<l}^{J-1} x_k x_l + O(\epsilon^3) \nonumber \\
        & = \sum_{k=0}^{J-1} x_k + \frac{1}{2}\left[ \left(\sum_{k=0}^{J-1} x_k \right)^2 - \sum_{k=0}^{J-1} x_k^2 \right] + O(\epsilon^3) \nonumber \\
        & =: B_1 + B_2 + O(\epsilon^3),
\end{align}
where the first and second order terms are evaluated to be
\begin{align}
    B_1 & = \frac{J(2n+J-1)}{2d}, \\
	B_2 & = \frac{J(J-1)(J[12n-7]+12n[n-1]+3J^2+2)}{24d^2},
\end{align}
and $\epsilon = n^2/d$.
Similarly, letting $y_k = [n-J-1+k]/d$,
\begin{align}
    \prod_{k=0}^J (1+[n-J-1+k]/d) & = \sum_{k=0}^J y_k + \frac{1}{2} \left[ \left(\sum_{k=0}^J y_k^2\right)^2 - \sum_{k=0}^J y_k^2 \right] + O(\epsilon^3) \nonumber \\
        & =: C_1 + C_2 + O(\epsilon^3),
\end{align}
with
\begin{align}
    C_1 & = \frac{(J+1)(2n-J-2)}{2d}, \\
    C_2 & = \frac{J(J+1)(12n^2 - 12n[J+2] + 3J^2 + 11J + 10)}{24d^2}.
\end{align}
We then have
\begin{align}
    \frac{d^B_J/p_J}{\binom{n+d-1}{n}^2} & = \left(1-\frac{1}{d}\right) (1+B_1+B_2) (1+C_1+C_2)^{-1} + O(\epsilon^3) \nonumber \\
        & = 1 + R_1 + R_2 + O(\epsilon^3), \\
    R_1 & = B_1 - C_1 - \frac{1}{d} \nonumber \\
        & = \frac{J(J+1)-n}{d}, \\
    R_2 & = B_2 - C_2 + C_1^2 - \frac{B_1}{d} + \frac{C_1}{d} - B_1 C_1 \nonumber \\
        & = \frac{2n^2 - 2n(2J[J+1]+1) + J(J+1)(J^2+J+2)}{2d^2}.
\end{align}
We now use this to compute the deficit in the change of entropy, as compared with the entropy for the informed observer:
\begin{align} \label{eqn:entropy_2nd_order}
    \DS{igno} - \DS{info} & = \sum_J p_J \ln\left( \frac{d_x^J/p_J}{\binom{n+d-1}{n}^2}\right) + p_J \ln p_J \nonumber \\
        & = \sum_J p_J \ln(1+R_1+R_2+O[\epsilon^3]) -H(\bm{p}) \nonumber \\
        & = \sum_J p_J \left(R_1 + R_2 - \frac{R_1^2}{2}\right) + O(\epsilon^3) -H(\bm{p}) ,
\end{align}
having used the expansion $\ln(1+x) = x - x^2/2 + \dots$ for small $x$.

In order to compute the first and second order terms in Eq.~\eqref{eqn:entropy_2nd_order} exactly, we need the following sums involving binomial coefficients:
\begin{align} \label{eqn:binom_sum1}
    \sum_{J=0}^n \binom{2n+1}{n+J+1} (2J+1) & = (2n+1) \binom{2n}{n},  \\ \label{eqn:binom_sum2}
    \sum_{J=0}^n \binom{2n+1}{n+J+1} (2J+1)J(J+1) & = (2n)(2n+1)\binom{2n-1}{n-1}.
\end{align}
These are both proved using the easily checked identity
\begin{align} \label{eqn:binom_diff}
    \frac{N-2k}{N}\binom{N}{k} & = \binom{N-1}{k} - \binom{N-1}{k-1} .
\end{align}
For Eq.~\eqref{eqn:binom_sum1}, we have (setting $k=n-J,\,N=2n+1$)
\begin{align}
    \sum_{J=0}^n \binom{2n+1}{n+J+1} (2J+1) & = \sum_{J=0}^n \binom{2n+1}{n-J} (2J+1) \nonumber \\
        & = \sum_{k=0}^n \binom{2n+1}{k} (2n+1-2k) \nonumber \\
        & = \sum_{k=0}^n (2n+1) \left[ \binom{2n}{k} - \binom{2n}{k-1} \right] \nonumber \\
        & = (2n+1) \binom{2n}{n}.
\end{align}
Similarly, for Eq.~\eqref{eqn:binom_sum2},
\begin{align}
    \sum_{J=0}^n \binom{2n+1}{n+J+1}(2J+1)J(J+1) & = \sum_{k=0}^n \binom{2n+1}{k} (2n+1-2k)(n-k)(n-k+1) \nonumber \\
        & = \sum_{k=0}^n (2n+1) \left[ \binom{2n}{k}-\binom{2n}{k-1} \right] (n-k)(n-k+1) \nonumber \\
        & = (2n+1) \sum_{k=0}^n \binom{2n}{k}(n-k)(n-k+1) - (2n+1)\sum_{k=0}^{n-1} \binom{2n}{k} (n-k-1)(n-k) \nonumber \\
        & = (2n+1) \sum_{k=0}^{n-1} \binom{2n}{k}(n-k) \left[ (n-k+1)-(n-k-1) \right] \nonumber \\
        & = (2n+1) \sum_{k=0}^{n-1} \binom{2n}{k}(2n-2k),
\end{align}
and by using Eq.~\eqref{eqn:binom_diff} with $N=2n$,
\begin{align}
    \sum_{J=0}^n \binom{2n+1}{n+J+1}(2J+1)J(J+1) & = (2n+1)(2n) \sum_{k=0}^{n-1} \binom{2n-1}{k}-\binom{2n-1}{k-1} \nonumber \\
        & = (2n+1)(2n) \binom{2n-1}{n-1}.
\end{align}
Recall that
\begin{equation}
    p_J = \frac{(n!)^2}{(2n+1)!} \binom{2n+1}{n+J+1} (2J+1),
\end{equation}
so the first order contribution is
\begin{align}
    \sum_{J=0}^n p_J R_1(J) & = \sum_{J=0}^n p_J \frac{J(J+1)-n}{d} \nonumber \\
        & = -\frac{n}{d} + \frac{(n!)^2}{d (2n+1)!} \binom{2n+1}{n+J+1} (2J+1)J(J+1) \nonumber \\
        & = -\frac{n}{d} + \frac{(n!)^2}{d (2n+1)!} (2n+1)(2n)\binom{2n-1}{n-1} \nonumber \\
        & = -\frac{n}{d} + \frac{(n!)^2(2n+1)(2n)(2n-1)!}{d(2n+1)!(n-1)!(n!)} \nonumber \\
        & = -\frac{n}{d} + \frac{n}{d} = 0.
\end{align}
The second order is
\begin{align}
    \sum_J p_J \left[R_2(J) - \frac{R_1(J)^2}{2} \right] & = \sum_{J=0}^n p_J \frac{n(n-2) - 2(n-1)J(J+1)}{2d^2} \nonumber \\
        & = \frac{n(n-2)}{2d^2} - \frac{2(n-1)}{2d^2} \sum_{J=0}^n p_J J(J+1) \nonumber \\
        & = \frac{n(n-2)}{2d^2} - \frac{2(n-1)}{2d^2} n \nonumber \\
        & = -\frac{n^2}{2d^2}.
\end{align}
Therefore, substituting the above into Eq.~\eqref{eqn:entropy_2nd_order}, we have
\begin{align}
     \DS{igno} - \DS{info} = -H(\bm{p}) - \frac{n^2}{2d^2} + O\left(\frac{n^3}{d^3}\right).
\end{align}

\subsection*{Fermions}
The method is the same as for bosons.
We expand $\frac{d^F_J/p_J}{\binom{d}{n}^2}$ to second order.
Letting $z_k = [k-n-J]/d$, we have
\begin{align}
    \prod_{k=1}^J (1 + [k-n-J]/d) & = F_1 + F_2 + O(\epsilon^3),
\end{align}
where
\begin{align}
    F_1 & = \sum_{k=1}^J z_k = \frac{-J(2n+J-1)}{2d}, \\
    F_2 & = \frac{1}{2}\left[ F_1^2 - \sum_{k=1}^J z_k^2 \right] = \frac{J(J-1)(2+3J^2+12n[n-1]+J[12n-7])}{24d^2}.
\end{align}
Similarly, letting $w_k = [k-n+1]/d$,
\begin{align}
    \prod_{k=0}^J (1+ [k-n+1]/d) & = G_1 + G_2 + O(\epsilon^3),
\end{align}
where
\begin{align}
    G_1 & = \sum_{k=0}^J w_k + \frac{(J+1)(J-2n+2)}{2d} , \\
    G_2 & = \frac{1}{2}\left[ G_1^2 - \sum_{k=0}^J w_k^2 \right] = \frac{J(J+1)(10+11J+3J^2-12n[J+2]+12n^2)}{24d^2}.
\end{align}
We then have
\begin{align}
    \frac{d^F_J/p_J}{\binom{d}{n}^2} & = \left(1+\frac{1}{d}\right)(1+F_1+F_2)(1+G_1+G_2)^{-1} + O(\epsilon^3) \nonumber \\
        & = 1 + T_1 + T_2 + O(\epsilon^3), \\
    T_1 & = F_1 - G_1 + \frac{1}{d} \nonumber \\
        & = \frac{-J(J+1)+n}{d} , \\
    T_2 & = F_2 - G_2 + G_1^2 + \frac{F_1}{d} - \frac{G_1}{d} - F_1 G_1 \nonumber \\
        & = \frac{2n^2 -2n(2J[J+1]+1) + J(J+1)(J^2+J+2)}{2d^2}.
\end{align}
Note that compared with the boson case, $T_1 = -R_1, \, T_2 = R_2$, thus the first order vanishes and we again have
\begin{align}
	\DS{igno} - \DS{info} = -H(\bm{p}) - \frac{n^2}{2d^2} + O\left(\frac{n^3}{d^3}\right).
\end{align}

\section*{Supplementary Note 6 -- Entropy $H(\mathbf{p})$ for large particle number} \label{app:entropy_large_n}
Here, we evaluate the entropy $H(\bm{p})$ for large particle number.
We take $n = m \gg 1$.
Starting from Eq.~\eqref{eqn:pj_cg_coeff}, we can rewrite
\begin{align}
	p_J & = (2J+1) \frac{(n!)^2}{(2n+1)!} \binom{2n+1}{n+J+1} \nonumber \\
		& = (2J+1) \frac{(n!)^2 2^{2n+1}}{(2n+1)!} b(n+J+1),
\end{align}
where $b(n+J+1) = 2^{-(2n+1)}\binom{2n+1}{n+J+1}$ follows a binomial distribution with $N+1$ trials and a success probability of $1/2$.

Using Stirling's approximation in the form $n! = \sqrt{2\pi}n^{n+1/2}e^{-n+O(1/n)}$~\cite{FlajoletSedgewick}, we have
\begin{align}
	\frac{(n!)^2}{(2n+1)!} & = \frac{n^{2n+1} e^{-2n + O(1/n)}}{\sqrt{2\pi} (2n+1)^{2n+3/2} e^{-2n-1+O(1/n)}} \nonumber \\
		& = (\sqrt{2\pi e}) \left(\frac{n}{2n+1}\right)^{2n+1} \frac{1}{(2n+1)^{1/2}} [1+O(1/n)] \nonumber \\
		& = \frac{\sqrt{2\pi}e}{2^{2n+1} \left(1+\frac{1}{2n}\right)^{2n+1} (2n+1)^{1/2}} [1+O(1/n)] \nonumber \\
		& = \frac{\sqrt{2\pi} e}{2^{2n+1} [e+O(1/n)] (2n+1)^{1/2}} [1+O(1/n)] \nonumber \\
		& = \frac{1}{2^{2n+1}} \sqrt{\frac{2\pi}{2n+1}} [1+ O(1/n)].
\end{align}
Using a local version of the central limit theorem~\cite[Chapter VII, Theorem 1]{Petrov1975Sums}, we can approximate $b(n+J+1)$ by a normal distribution with mean $(2n+1)/2$ and variance $(2n+1)/4$, obtaining
\begin{align}
	p_J & = (2J+1) \sqrt{\frac{2\pi}{2n+1}}[1+O(1/n)] \left[ \frac{e^{ -\frac{(J+1/2)^2}{n+1/2} }}{\sqrt{2\pi(2n+1)/4}} + o(n^{-1/2}) \right] \nonumber \\
		& = (2J+1) \left[ \frac{e^{-\frac{(J+1/2)^2}{n+1/2} }}{n+1/2} + o(1/n) \right] [1+ O(1/n)] \nonumber \\
		& = (2J+1) \frac{e^{-\frac{(J+1/2)^2}{n+1/2} }}{n+1/2} [1 + o(1)] [1+ O(1/n)] \nonumber \\
		& = (2J+1) \frac{e^{-\frac{(J+1/2)^2}{n+1/2} }}{n+1/2} [1 + o(1)],
\end{align}
where $o(f)$ denotes an error term going to zero strictly faster than $f$.
Then 
\begin{align}
	\ln p_J & = \ln (2J+1) - \ln(n+1/2) - \frac{(J+1/2)^2}{n+1/2} + o(1),
\end{align}
so the entropy is approximated by
\begin{align} \label{eqn:entropy_mid_approx}
	H(\bm{p}) & = - \sum_{J=0}^n p_J \left[ \ln (2J+1) - \ln(n+1/2) - \frac{(J+1/2)^2}{n+1/2} + o(1) \right] \nonumber \\
		& = \ln(n+1/2) + o(1) + [1 + o(1)] \sum_{J=0}^n (2J+1) \frac{e^{-\frac{(J+1/2)^2}{n+1/2} }}{n+1/2} \left[ -\ln(2J+1) + \frac{(J+1/2)^2}{n+1/2} \right].
\end{align}
For large $n$, we expect that the sum can be approximated by an integral.
To show this, we can use the simplest version of the Euler-Maclaurin formula:
\begin{align}
	\sum_{J=0}^n f(J) & = \int_0^n f(x) \, \dd x + \int_0^n \left( x- \lfloor x \rfloor - \frac{1}{2} \right) f'(x) \, \dd x + \frac{f(0) + f(n)}{2}, \nonumber \\
	f(x) & := (2x+1) \frac{e^{-\frac{(x+1/2)^2}{n+1/2}} }{n+1/2} \left[ -\ln(2x+1) + \frac{(x+1/2)^2}{n+1/2} \right].
\end{align}
Firstly, we have
\begin{align}
	f(0) & = \frac{e^{-\frac{1}{4(n+1/2)}}}{n+1/2} \cdot \frac{1}{4(n+1/2)} = O(n^{-2}) , \nonumber \\
	f(n) & = 2 e^{-(n+1/2)} \left[ -\ln(2n+1) + (n+1/2) \right] = O(n e^{-n}).
\end{align}
Along these lines, it is not hard to see that shifting the initial point from $x=0$ to $x=1/2$ leads to an $o(1)$ error, so we change variables to $y = x+1/2$ and let $g(y) := f(y-1/2)$.
Additionally, the upper limit can be extended to infinity with an error which can be verified to be $O(e^{-n}\mathrm{poly}[n,\ln n])$.
For the remainder integral, we let $k=(n+1/2)^{-1}$ and use
\begin{align}
	g(y) & = 2k e^{-ky^2} \left[-y\ln(2y) + ky^3 \right], \nonumber \\
	g'(y) & = 2k e^{-ky^2} \left[2k y^2 \ln(2y) -2k^2y^4 - \ln(2y) -1 + 3k y^2 \right].
\end{align}
Together with $\abs{y - \lfloor y \rfloor - 1/2 } \leq 1/2$, we have
\begin{align}
	\abs{ \int_0^\infty \left(y-\lfloor y \rfloor - \frac{1}{2}\right) g'(y) \, \dd y } & \leq \abs{\int_0^\infty 2k^3 y \ln(2y) e^{-ky^2}\, \dd y} + \abs{\int_0^\infty 2k^3 y^4 e^{-ky^2} \, \dd y} + \abs{\int_0^\infty k \ln(2y) e^{-ky^2} \, \dd y} \nonumber \\
		& \quad + \abs{ \int_0^\infty k e^{-ky^2} \, \dd y } + \abs{\int_0^\infty 3k^2 y^2 e^{-ky^2} \, \dd y} \,
\end{align}
in which the individual integrals can be evaluated with the highest order being $O\left(\frac{\ln n}{n}\right) = o(1)$.

Overall, therefore,
\begin{align}
	\sum_{J=0}^n f(J) & = \int_0^\infty g(y) \, \dd y + o(1) \nonumber \\
		& = \frac{1}{2}\left(\ln k + \gamma \right) - \ln 2 + 1 + o(1) \nonumber \\
		& = -\frac{1}{2}\ln n + \frac{\gamma}{2} - \ln 2 + 1 + o(1),
\end{align}
where $\gamma = 0.557\dots$ is the Euler-Mascheroni constant.
Putting this into Eq.~\eqref{eqn:entropy_mid_approx},
\begin{align}
	H(\bm{p}) & = \frac{1}{2}\ln n + \frac{\gamma}{2} - \ln 2 + 1 + o(1) \nonumber \\
		& = \frac{1}{2}\ln n + 0.595... + o(1).
\end{align}

\end{document}